\documentclass[a4paper,fleqn,10pt]{article}
\usepackage{cite,mcite}
\pdfoutput=1
\input{global}

\hypersetup{
  pdfauthor={Max Knobbe, Frank Krauss, Daniel Reichelt, Steffen Schumann}
  pdftitle={Measuring Higgs Branching Ratios at Future Lepton Colliders},
  pdfkeywords={Higgs Boson, Lepton Colliders, Event Shapes, QCD, Event Generators}
}
\begin{document}
\preprint{ IPPP/23/26 \\ MCNET-23-06 }
\title{Measuring Hadronic Higgs Boson Branching Ratios at Future Lepton Colliders}
\author{Max Knobbe$^1$, Frank~Krauss$^2$, Daniel Reichelt$^2$, Steffen Schumann$^1$}
\institute{
 $^1$ Institut f\"ur Theoretische Physik, Georg-August-Universit\"at G\"ottingen, G\"ottingen, Germany\\
 $^2$ Institute for Particle Physics Phenomenology, Durham University, Durham DH1 3LE, UK
}

\newcommand{\dr}[1]{\textbf{\textcolor[HTML]{3B87CF}{#1 --dr}}}
\newcommand{\fk}[1]{\textcolor[HTML]{CF3B87}{#1 --fk}}
\newcommand{\sts}[1]{\textcolor[HTML]{00FFFF}{#1 --sts}}

\maketitle
\begin{abstract}
We present a novel strategy for the simultaneous measurement of Higgs-boson branching ratios into gluons and light quarks at a future lepton collider operating in the Higgs-factory mode.
Our method is based on template fits to global event-shape observables, and in particular fractional energy correlations, thereby exploiting differences in the QCD radiation patterns of quarks and gluons. In a constrained fit of the deviations of the light-flavour hadronic Higgs-boson branching ratios from their Standard Model expectations, based on an integrated luminosity of $5\,\text{ab}^{-1}$, we obtain $68\%$ confidence level limits of $\mu_{gg}=1 \pm 0.05$ and $\mu_{q\bar{q}}<21$.
\end{abstract}

\section{Introduction}
\label{Sec::intro}

Electron--positron colliders operating as Higgs-boson factories constitute one of the main options for future accelerator-based high-energy experiments~\cite{FCC:2018byv,FCC:2018evy}.
Similar to the \LEP experiments of the 1990's and their precision determination of the $Z$-boson properties, the primary target of such a facility will be the profound analysis of the Higgs-boson properties.
This encompasses in particular measuring its width, branching ratios, and kinematic distributions of its decay products with unrivalled precision, searching for minute deviations from Standard Model predictions, usually formulated in the language of Effective Field Theories, see, e.g.~\cite{Weinberg:1979sa,Buchmuller:1985jz,Grzadkowski:2010es}.
One of the observables that attracted considerable attention is the branching ratio of the Higgs boson, and hence its coupling, to gluons, which at a hadron collider such as the \LHC can be accessed only through its (dominant) gluon-fusion production mode.

Due to the much cleaner final states produced in lepton--lepton annihilation, at a future $e^+e^-$ collider this decay can be studied directly in events where the Higgs boson decays into jets.
Ignoring for the moment decays into intermediate $W$ and $Z$-boson pairs, the gluon decay mode can be extracted by vetoing QCD events with displaced vertices which emerge in the weak decays of $b$ and $c$ hadrons~\cite{Azzi:2012yn}, a technique also underpinning measurements of branching ratios to heavy quarks at the LHC~\cite{CMS:2018nsn,ATLAS:2020jwz,CMS:2019hve,ATLAS:2022ers}.
Having thus eliminated the two main competitor decay modes with similar characteristics, the remaining QCD events are due to the Higgs boson coupling to the light degrees of freedom.
Realising that the Yukawa coupling to the $u$, $d$, and $s$ quarks is negligible in the Standard Model yields the  coupling of the Higgs boson to gluons.
Results based on this strategy indicate branching ratio measurements at relative per-cent accuracy~\cite{deBlas:2019rxi,Walker:2022yml}.

In this study we will advocate an alternative approach to this measurement, that is agnostic towards the presence of displaced vertices and is underpinned by fits to event-shape observables alone. This is in line with previous ideas to use event- or jet shapes as theoretically well understood taggers \cite{Amoroso:2020lgh, Caletti:2021ysv}, but goes a step further by never explicitly assigning a flavour to a given event.
Such a strategy relies on two well-established properties of QCD radiation patterns, namely firstly that gluons carry two colours resulting in a ratio $C_A/C_F=9/4$ of colour charges with respect to quarks and hence about twice as many emissions, and secondly that the finite masses of the heavy quarks shield the collinear divergence of gluon emission, thereby depleting their QCD radiation in this region, a phenomenon also known as the "dead cone" effect~\cite{Dokshitzer:1991fd}.
Combining both effects allows the placement of direct constraints to the sum of the light-quark Yukawa couplings. 

In particular we will use fractional energy correlations~\cite{Banfi:2004yd} which are geared towards a systematic study of the collinear regions of the radiation pattern.
Our studies here supplement the existing strategy for the measurement of the Higgs-boson branching ratio to gluons, see also Ref.~\cite{Wang:2023azz} for a recent study using jet charge as a discriminating variable. 
They furthermore provide an alternative to first attempts to measure the Yukawa coupling to light quarks through rare decays such as $H\to \phi\gamma$ at the \LHC~\cite{Perez:2015lra,ATLAS:2017gko}. 

In relying on event-shape variables alone, hadronisation effects constitute the dominant systematic uncertainty when fitting Monte Carlo results to (synthetic) data.
To account for this we re-tune the new cluster-hadronisation model~\cite{Chahal:2022rid} of the \Sherpa event generator~\cite{Sherpa:2019gpd} to \LEP data, and quantify the resulting uncertainties through repeated tunes with varying input data. We refer to the resulting sets of alternative hadronisation parameter values as \emph{replica tunes}.

Our discussion is structured as follows: In Section 2 we detail the setup of our analysis, and in particular the event-shape observables we use.
This is followed by Section 3 where we describe our simulations with \Sherpa and we put special emphasis on the re-tuning of its fragmentation model.
In Section 4 we discuss the results emerging from fits to various event-shape distributions, with and without soft-drop grooming the hadronic final state, which we present as allowed two-dimensional regions of values for the deviations $\mu_{gg}$ and $\mu_{qq}$ of the Higgs boson branching ratios into gluons and light quarks.
We conclude and summarise our study in Section 5.

\section{Analysis of event shapes in $e^+e^-\to ZH$}\label{sec:analysis}

We here propose an analysis at a future lepton collider operating at the working point for Higgs-strahlung production, \emph{i.e.}\ $\sqrt{s}\gtrsim m_H+m_Z$.
Our goal is to consider hadronic Higgs-boson decays, where we separate the branchings to gluons, light (up, down and strange) quarks, charm quarks and bottom quarks.
Experimentally, those will at first all be seen as hadronic channels.
However, due to the difference in the QCD radiation pattern between quarks and gluons, and the imprints of finite quark masses, one can expect differences in observables such as the well-studied event shapes.

Following the selection cuts of Ref.~\cite{Azzi:2012yn}, we identify $Z$-boson candidates as pairs of opposite-sign leptons within $\pm 5~\text{GeV}$ of the nominal $Z$ mass.
The reconstructed $Z$-boson is required to have at least a transverse  momentum of $p_{T,Z} > 10~\text{GeV}$ and a longitudinal momentum of at most $50~\text{GeV}$.
To suppress irreducible backgrounds from $ZZ$ events, we require for the opening angle between the two leptons $\theta_{l^+l^-} <100^\circ$.
We additionally ask for a total hadronic mass of all other particles to be at least $m_\text{had} > 75~\text{GeV}$.
In order to select events where the hadronic final state is likely to originate from a decaying Higgs-boson, we constrain the recoil mass of the lepton pair, defined as
\begin{equation}
    m_\text{recoil}^2 = s + m_Z^2 - 2 \sqrt{s} (E_{l^+}+E_{l^-})~,
\end{equation}
see also~\cite{FCC:2018evy}, to be similar to the Higgs-boson mass.
In practice we use $120\,\text{GeV} < m_\text{recoil} < 130\,\text{GeV}$.

We base the calculation of event-shape observables on charged-particle tracks\footnote{Note, the restriction to charged tracks is not strictly needed for our particle-level analysis, but we expect that the experimental resolution of angular separations improves when using charged-particle tracking information.} and consider the family of fractional energy correlations~\cite{Banfi:2004yd}
\begin{equation}\label{eq:frac_en_corr}
    \text{FC}_{x} \equiv \sum\limits_{i\neq j}\frac{E_iE_j|\sin\theta_{ij}|^x(1-|\cos\theta_{ij}|)^{1-x}}{(\sum_iE_i)^2}\Theta\left[(\vec{q}_i\cdot\vec{n}_T)(\vec{q}_j\cdot\vec{n}_T)\right]~,
\end{equation}
with $x=0.5,1,1.5$.
The sums run over all charged tracks $i$ and $j$ with respective energies $E_{i, j}$ and three-momenta $\vec{q}_{i, j}$.
All energies and angles are evaluated in the Higgs-boson rest frame, which we reconstruct as the full charged final state, excluding the two leptons from the $Z$-boson decay.
We can analyse the behaviour of this class of observables in response to a single soft-gluon emission off a hard parton from the hadronic decay as commonly done in the context of resummation calculations \cite{Banfi:2004yd}. In terms of the soft-gluon transverse momentum $k_t$ and rapidity $\eta$ relative to the hard parton, the fractional energy observables scale like
\begin{equation}\label{eq:scaling}
    \text{FC}_x \sim \frac{k_t}{Q}e^{b\eta}\,,\;\;\text{with}\;\; b=1-x\,.
\end{equation}

Hence $x=1$ corresponds to a purely transverse-momentum like scaling; larger (smaller) values of $x$ will give a higher weight for pairs of particles with smaller (larger) opening angles. The fractional energy correlations are very similar to jet angularities that are studied at the \LHC in the context of quark and gluon tagging, see for example~\cite{ATLAS:2019kwg,CMS:2021iwu,ALICE:2021njq,Caletti:2021oor,Reichelt:2021svh}.
The choice $x=1.5$ corresponds to the Les-Houches angularity~\cite{Gras:2017jty}.
By the Heaviside function in Eq.~\eqref{eq:frac_en_corr}, the fractional energy correlations are implemented as sum over the contributions from two hemispheres defined by the axis $\vec{n}_T$.
We follow the original definition and use the thrust variable to define the reference axis, and hence the hemispheres.
There are further standard observables that are written in this way, for example total hemisphere broadening,
\begin{equation}
    B_\text{tot}=B_++B_-~,\;\;B_{\pm} = \frac{\sum_i |\vec{q}_i\times \vec{n}_T|\Theta\left[\pm\vec{q}_i\cdot\vec{n}_T\right]}{\sum_i |\vec{q}_i|}~,
\end{equation}
or total mass
\begin{equation}
    m_\text{tot}^2 = m_+^2 + m_-^2~,\;\;m_\pm^2 = \frac{\left(\sum_i \vec{q}_i\Theta\left[\pm\vec{q}_i\cdot\vec{n}_T\right]\right)^2}{\sum_i |\vec{q}_i|}~.
\end{equation}
We also considered properties of individual hemispheres, for example the mass of the heavier hemisphere and the broadening of the wider one, but did not observe any noteworthy increase in performance and hence focus on the ones described above. In terms of Eq.~\eqref{eq:scaling}, the broadening scales with $b=0$ and behaves similar to $\text{FC}_{1}$ while the mass would correspond to $\text{FC}_0$, \emph{i.e.}\ $b=1$, which we do not analyse here.

As an additional handle, we employ soft-drop grooming~\cite{Larkoski:2014wba} with the goal to reduce hadronisation corrections.
While the soft-drop grooming algorithm was originally developed to mitigate the contamination of jets from effects that are typically simulated as underlying event or multiple parton interactions, it has been shown to be effective in mitigating non-perturbative corrections to event-shape observables in leptonic and hadronic collisions as well~\cite{Baron:2018nfz,Marzani:2019evv,Baron:2020xoi}.
We apply the algorithm individually to the two event hemispheres.
To this end, we recluster their respective constituents using the Cambridge/Aachen jet algorithm~\cite{Dokshitzer:1997in,Wobisch:1998wt}, then undoing the last clustering step between the subjets $i,j$ and checking for the soft-drop condition
\begin{equation}\label{eq:sd_cond}
    \frac{\min{\left[E_i,E_j\right]}}{E_i+E_j} > z_\text{cut}~.
\end{equation}
If this condition is satisfied, the procedure terminates.
Otherwise, the softer of the subjets, with smaller energy, is discarded and the procedure is repeated for the harder one.
This continues until Eq.~\eqref{eq:sd_cond} is true, or the remaining subjet consists of only one track.
Note that other references include an angular dependence in the soft-drop condition, whereas we here only consider the $\beta=0$ case of~\cite{Larkoski:2014wba}, which is equivalent to the modified mass drop tagger~\cite{Butterworth:2008iy,Dasgupta:2013ihk}.
We also restrict our study to the conventional case $z_\text{cut}=0.1$.
The observables are then calculated on the remaining particles after grooming, however, normalised by the hadronic energy before grooming.
This treatment is necessary in order to define collinear safe observables~\cite{Marzani:2019evv}.

For our final results, histograms for the differential distribution of event shapes $v$ are constructed as sums over the individual decay channels
\begin{equation}\label{eq:histo_sum}
    \frac{\mathrm{d}\sigma}{\mathrm{d}v} = \sum_{i\in\{q\bar{q},c\bar{c},b\bar{b},gg,WW,ZZ\}} \mu_i \frac{\mathrm{d}\sigma_i}{\mathrm{d}v} + \frac{\mathrm{d}\sigma_{ZZ}}{\mathrm{d}v}~,
\end{equation}
where the sum runs over the hadronic decay modes of the Higgs-boson, into light ($q\bar{q}$), charm ($c\bar{c}$) and bottom ($b\bar{b}$) quarks, gluons ($gg$), as well as two hadronically decaying vector bosons.
In the last term we add the irreducible background from $ZZ$ production.
The factors $\mu_i$ parametrise deviations from the Standard Model (SM) partial Higgs-boson decay widths, with the SM corresponding to $\mu_i = 1$ $\forall\,i$.
Ultimately, we aim for an experimental determination of the coefficients $\mu_i$.
In the following, we explore the possibility to set limits on simultaneous deviations of $\mu_{gg}$, $\mu_{q\bar{q}}$ and $\mu_{b\bar{b}}$ from $1$, while leaving the total cross section unchanged.
To achieve this, we scan different points in $\mu_{gg}$ and $\mu_{q\bar{q}}$, fixing $\mu_{c\bar{c}}=\mu_{WW} = \mu_{ZZ} = 1$, and imposing the constraint
\begin{equation}\label{eq:mu_const}
    \mu_{b\bar{b}} = 1 - (\mu_{gg}-1) \frac{\sigma_{gg}}{\sigma_{b\bar{b}}} - (\mu_{q\bar{q}}-1)\frac{\sigma_{q\bar{q}}}{\sigma_{b\bar{b}}}~.
\end{equation}

\section{Simulation with \sherpa}

To simulate particle-level events we use the \sherpa event generator.
The main physics aspects of the framework are documented in \cite{Sherpa:2019gpd}, while we here work with a pre-release version 3.0$\beta$~\cite{Sherpa3.0.beta}.
We use \sherpa's default dipole shower based on Catani--Seymour factorisation~\cite{Schumann:2007mg}, that supports
finite parton masses in the splitting kernels and branching kinematics.
Parton-to-hadron transitions are described by \sherpa's built-in cluster-hadronisation model~\cite{Chahal:2022rid} and hadron decays are treated by its internal decay package~\cite{Sherpa:2019gpd,Gleisberg:2008ta}.
We will comment below on a dedicated hadronisation-parameter tune based on sensitive measurements from \LEP experiments.
We analyse our simulated data with the \rivet  package~\cite{Bierlich:2019rhm} and use the \Contur tool~\cite{Butterworth:2016sqg,Buckley:2021neu} for statistical analyses and the calculation of exclusion limits.

\subsection{FCC-ee setup}
We assume the operating conditions for a Future Circular Collider, FCC-ee,  running at a centre-of-mass energy of $\sqrt{s}=240~\text{GeV}$~\cite{FCC:2018byv,FCC:2018evy}.
We simulate the processes $e^+e^-\to Z(\to\mu^+\mu^-) H(\to q\bar{q})$ at order $\alpha_\text{EW}^3y^2_q$ separately for $q=u,d,s$, collectively referred to as light-quark decays, $q=c$ and $q=b$, and $e^+e^-\to Z(\to\mu^+\mu^-) H(\to gg)$.
We here assume the one-loop decay $H\to gg$ in the heavy top-quark limit, treating it through an effective $ggH$ vertex~\cite{Ellis:1975ap,Shifman:1979eb,Kniehl:1995tn}.
We take into account the Higgs-boson decays to $WW^*$ and $ZZ^*$ by generating a sample of $e^+e^-\to Z(\to\mu^+\mu^-) H$ events where the Higgs is forced to decay into $Wq\bar{q}'$ or $Zq\bar{q}$ and the on-shell vector bosons likewise decay into quarks. We rescale our leading-order results to the branching ratios from~\cite{LHCHiggsCrossSectionWorkingGroup:2016ypw}, corresponding to 
\begin{eqnarray}
    &&\mbox{\rm BR}(H\to b\bar{b}) = 56.81 \%\;,\;\;\;
    \mbox{\rm BR}(H\to c\bar{c}) = 2.82 \%\;,\;\;\; \mbox{\rm BR}(H\to gg) = 8.112 \%\;,
\end{eqnarray}
by appropriately adjusting the event weights in the final samples.  
While we neglect contributions from $q=u,d$, we estimate the $H\to s\bar{s}$ branching ratio by
scaling the $H\to c\bar{c}$ result~\cite{Albert:2022mpk}, \emph{i.e.}\
\begin{eqnarray}
    \sum_{q=u,d,s,}\mbox{\rm BR}(H\to q\bar q)\approx \mbox{\rm BR}(H\to s\bar s) \approx \left(\frac{m_s}{m_c}\right)^2 \mbox{\rm BR}(H\to c\bar{c}) = \left(11.72\right)^{-2} \mbox{\rm BR}(H\to c\bar{c})\;.\;\; 
\end{eqnarray}

Note that we handle $c$ and $b$ quarks as massive in the parton-shower evolution~\cite{Schumann:2007mg}.
Finally, we simulate a sample for resonant and non-resonant di-boson production, {\it i.e.}\ $e^+e^-\to \mu^+\mu^-q\bar{q}$ at order $\alpha_\text{EW}^4$, which we refer to as $ZZ$ background.  We do not include any higher-order corrections at this stage of the simulation.

The computed cross sections are scaled to an integrated luminosity of ${\mathcal{L}}=5~\text{ab}^{-1}$ to account for the full projected statistics accumulated by the FCC-ee at this centre-of-mass energy, and we include another factor of two to emulate the use of both electron and muon channels.
We construct histograms for the considered set of observables according to Eq.~\eqref{eq:histo_sum}.
The statistical errors are scaled accordingly with the estimated number of entries for a given bin $i$, $N_i$, as $\sqrt{N_i}$.
This is combined with a covariance matrix for the systematic variations of hadronisation-model parameters, see below, derived as
\begin{equation}
    V_{ij} = \langle N_i N_j \rangle - \langle N_i\rangle\langle N_j\rangle~,
\end{equation}
where the average is taken over runs with different tuning parameters.

\begin{figure}
  \includegraphics[width=.49\textwidth]{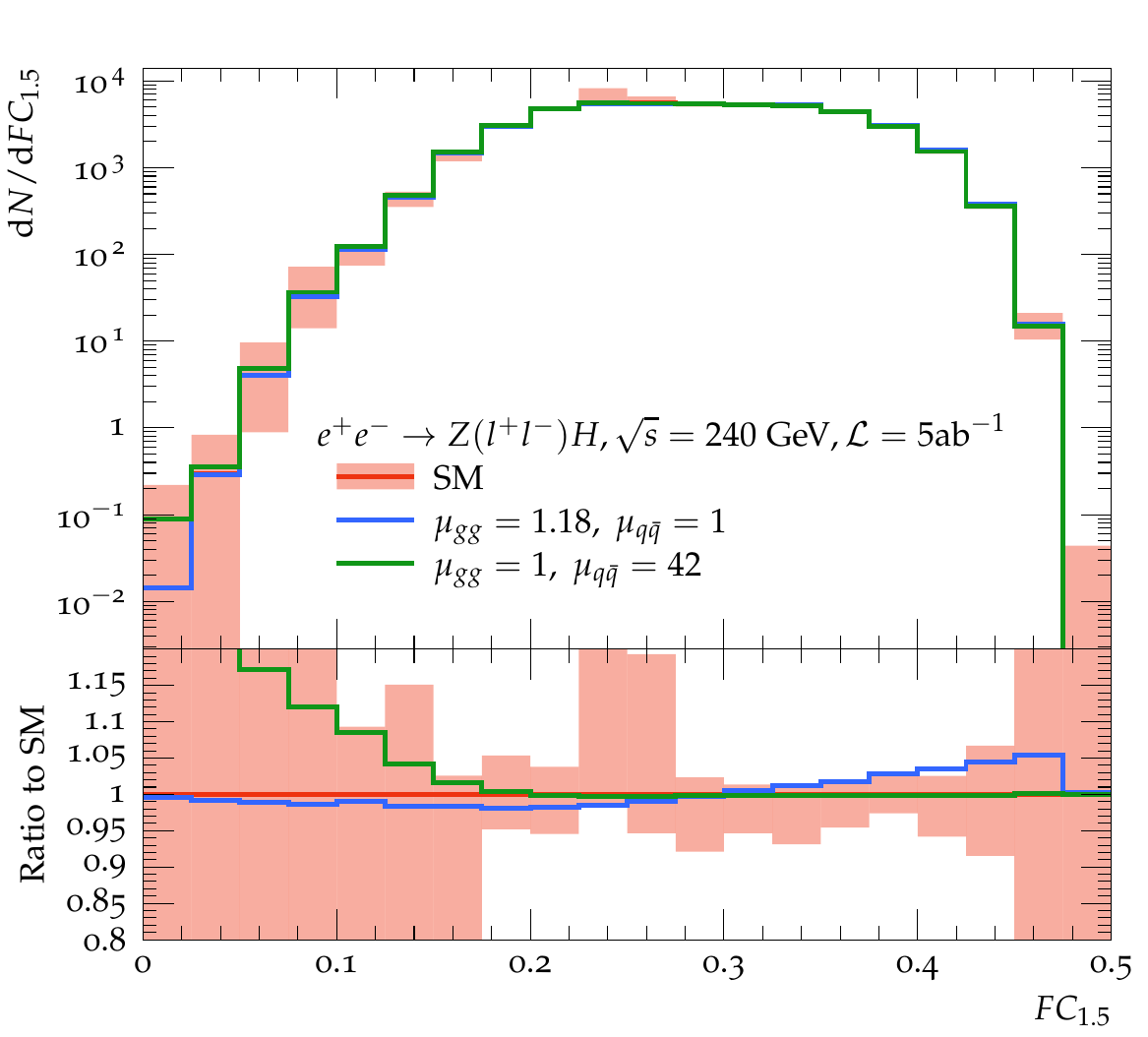}
  \includegraphics[width=.49\textwidth]{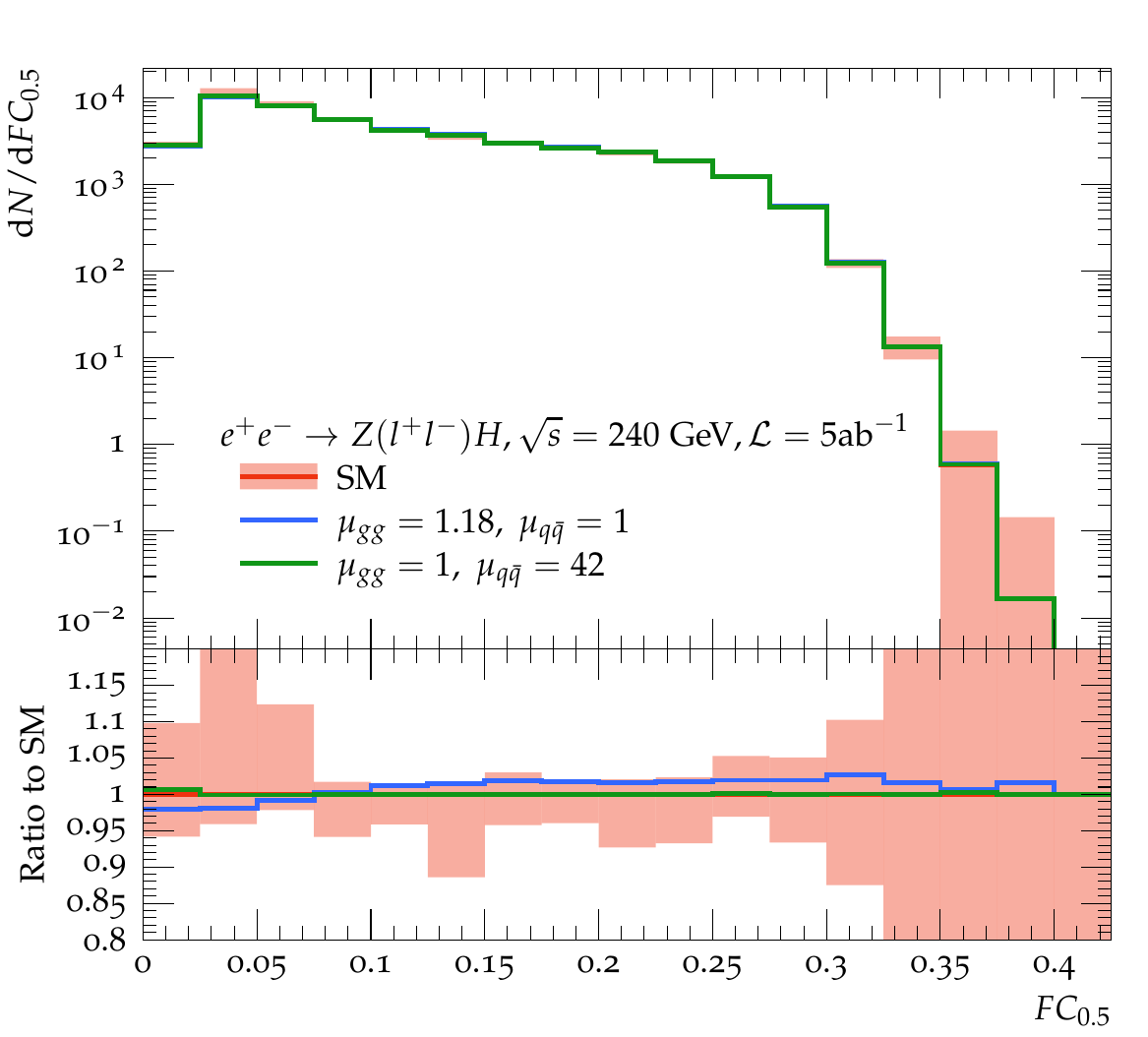}
  \caption{Predictions for fractional energy correlations $FC_{1.5}$ (left) and $FC_{0.5}$ (right) at the FCC-ee for the SM and two hypotheses for modified couplings of the Higgs-boson to QCD partons. The bands indicate the combined statistical and systematic errors of the SM prediction.}
    \label{fig:fcx_plots}
\end{figure}

In Fig.~\ref{fig:fcx_plots} we show example predictions for the fractional energy correlations $FC_{1.5}$ and $FC_{0.5}$.
We illustrate the SM distribution obtained from our simulations with \Sherpa as well as the variations corresponding to two representative sampling points in the $(\mu_{gg},\mu_{q\bar{q}})$ plane, {\emph{i.e.}}\ $\{\mu_{gg}=1,\mu_{q\bar{q}}=4\}$ and $\{\mu_{gg}=1.18,\mu_{q\bar{q}}=1\}$.

\subsection{\LEPOne setup and Tuning}
For the present study we perform dedicated tunes of \sherpa's new cluster-fragmentation model \Ahadic~\cite{Chahal:2022rid}, focusing on event-shape observables.
We provide non-perturbative (tuning) uncertainties through replica tunes.

Similar to previous tunes, we concentrate on observables for hadronic final states in electron--positron annihilation accurately measured at \LEPOne.
Our simulations rely on next-to-leading order (NLO) QCD matrix elements for $e^+e^-\to q\bar q+\{0,1\}j$ dressed by the parton shower, using the \sherpa implementation of the \MEPSatNLO formalism~\cite{Hoeche:2012yf}.
The contributing tree-level amplitudes are obtained from the
built-in matrix element generators \Comix~\cite{Gleisberg:2008fv} and \Amegic~\cite{Krauss:2001iv}, while the required one-loop amplitudes are obtained from \OpenLoops~\cite{Buccioni:2019sur}.
For the hard-scattering amplitudes we consider $q=u,d,s,c$ massless and only take into account mass-effects for $q=b$, therefore adding explicitly the tree-level contribution for the 4$b$ final state.
As jet-separation parameter in the merging prescription we use $\log_{10}\left(Q^2_{\text{cut}}/s\right)=-2$~\cite{Hoeche:2009rj}.

We employ the \Apprentice tuning tool~\cite{Krishnamoorthy:2021nwv} in combination with analyses provided by \rivet~\cite{Bierlich:2019rhm}.
We simultaneously tune all parameters listed in Table~\ref{tab:tuning_results} in Appendix~\ref{app:tuning_details}, starting from rather wide parameter ranges, decreasing the intervals in a sequence of tunes.
\Apprentice uses actual generator runs with varied hadronisation parameters to construct a bin-wise polynomial surrogate of the Monte-Carlo response for observables of interest.
To narrow down the tuning ranges in subsequent iterations, we construct multiple such surrogates each with a different set of generator runs and in this way find equivalent tunes, with results of similar quality.
The outcome of such set of tunes is then used to shrink the parameter ranges for the next iteration.
After the final iteration, the obtained family of equivalent tunes is used to estimate the remaining non-perturbative uncertainties by interpreting its members as replica tunes, \emph{i.e.}\ by re-running the Monte-Carlo simulation with the alternative parameter values.

Our observable selection for the new generator tunes is similar to the ones used for the initial \Ahadic tunes presented in \cite{Chahal:2022rid}, as well as its predecessor~\cite{Winter:2003tt}.
It consists mostly of mean and differential charged-particle multiplicities~\cite{ALEPH:1991ldi}, event shapes like thrust and its minor and major variants~\cite{ALEPH:2003obs,DELPHI:1996sen}, as well as the $b$-quark fragmentation function~\cite{ALEPH:2001pfo,OPAL:2002plk}.
Furthermore, we consider jet-rates for the Durham algorithm~\cite{JADE:1999zar}, and a selection of multiplicities of identified hadrons~\cite{ParticleDataGroup:2008zun}, thereby aiming for a general tune suitable for event-shape and jet observables. The complete list of analyses and differential distributions can be found in Table~\ref{tab:tuning_analyses} of Appendix~\ref{app:tuning_details}.

Figure~\ref{fig:tuning_plots} shows exemplary results for observables used in the tuning, including the final \sherpa prediction and the corresponding non-perturbative tune uncertainty indicated by the light-blue band.
To allow for sufficient freedom in the variations for all 16 parameters considered in the tuning, we provide 50 replica tunes.
They represent our tuning uncertainties by having each replica tuned with a different, random subset of Monte-Carlo runs.
We extract the uncertainty bands in Fig.~\ref{fig:tuning_plots} by re-running the simulation for each of the replica tunes, and plot the envelope of the resulting deviations.
We find good agreement between our \sherpa predictions and data, with deviations of the central tune to the data being largely covered by our estimated non-perturbative uncertainties.

\begin{figure}
  \includegraphics[width=.33\textwidth]{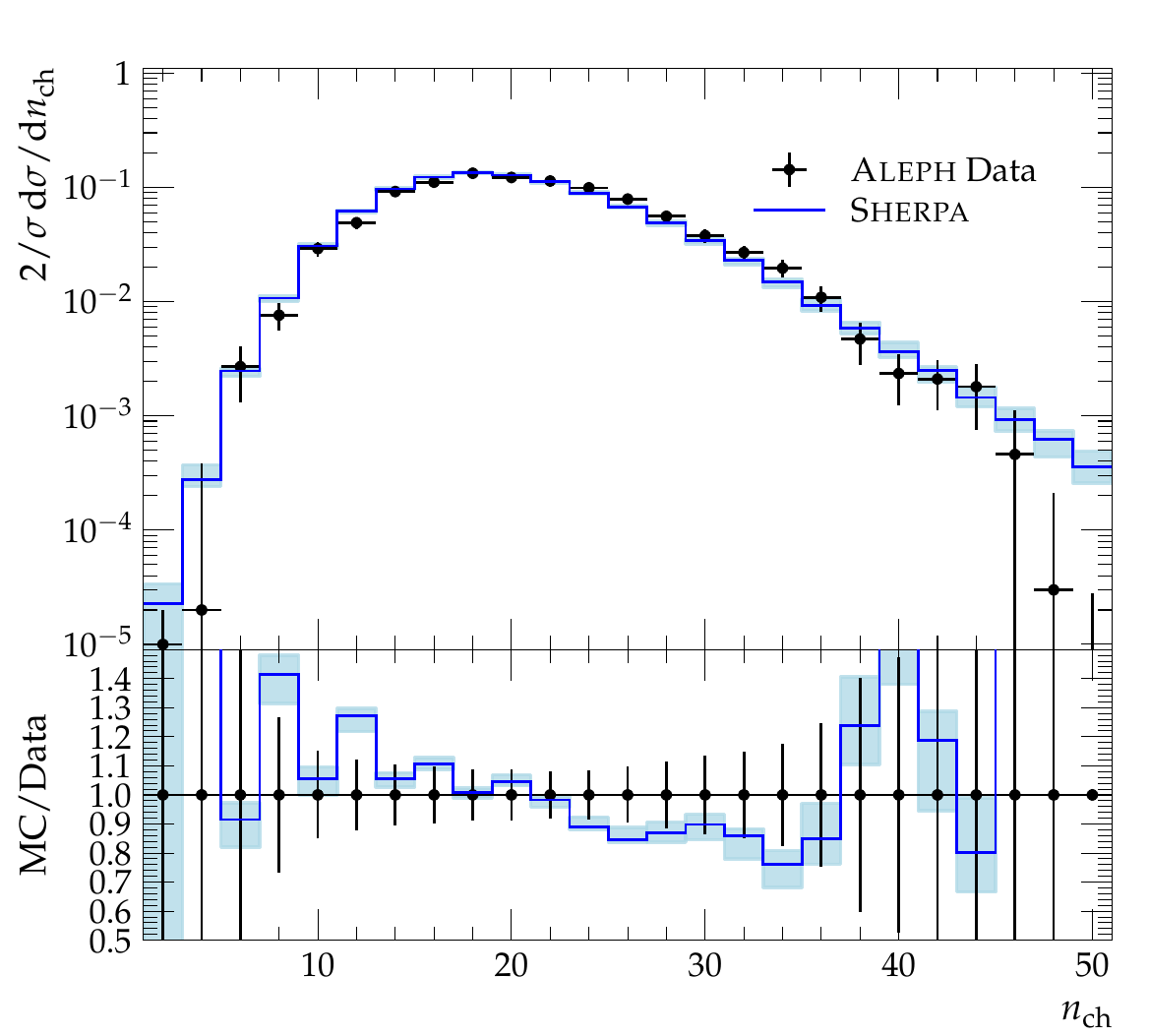}
  \includegraphics[width=.33\textwidth]{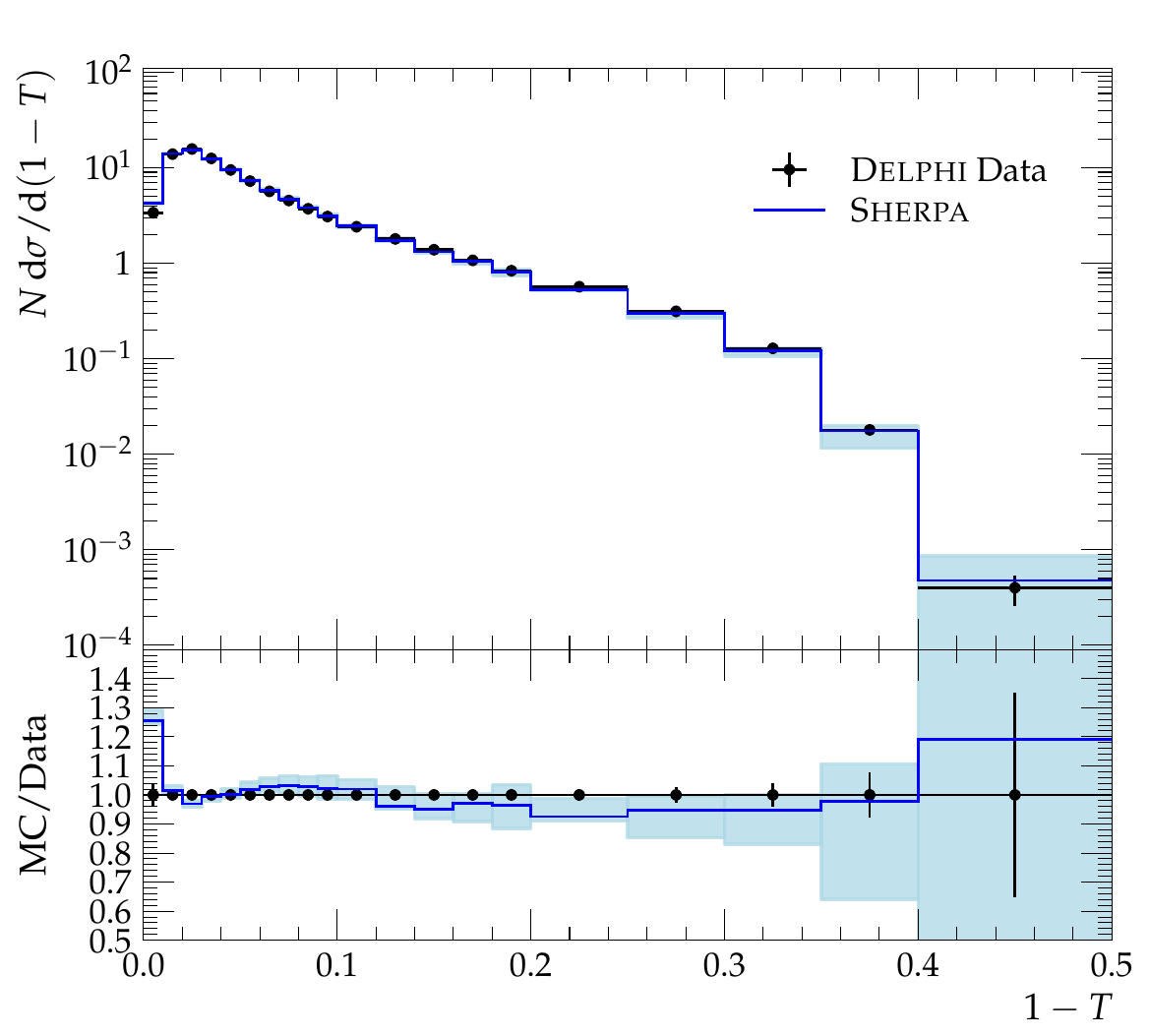}
  \includegraphics[width=.33\textwidth]{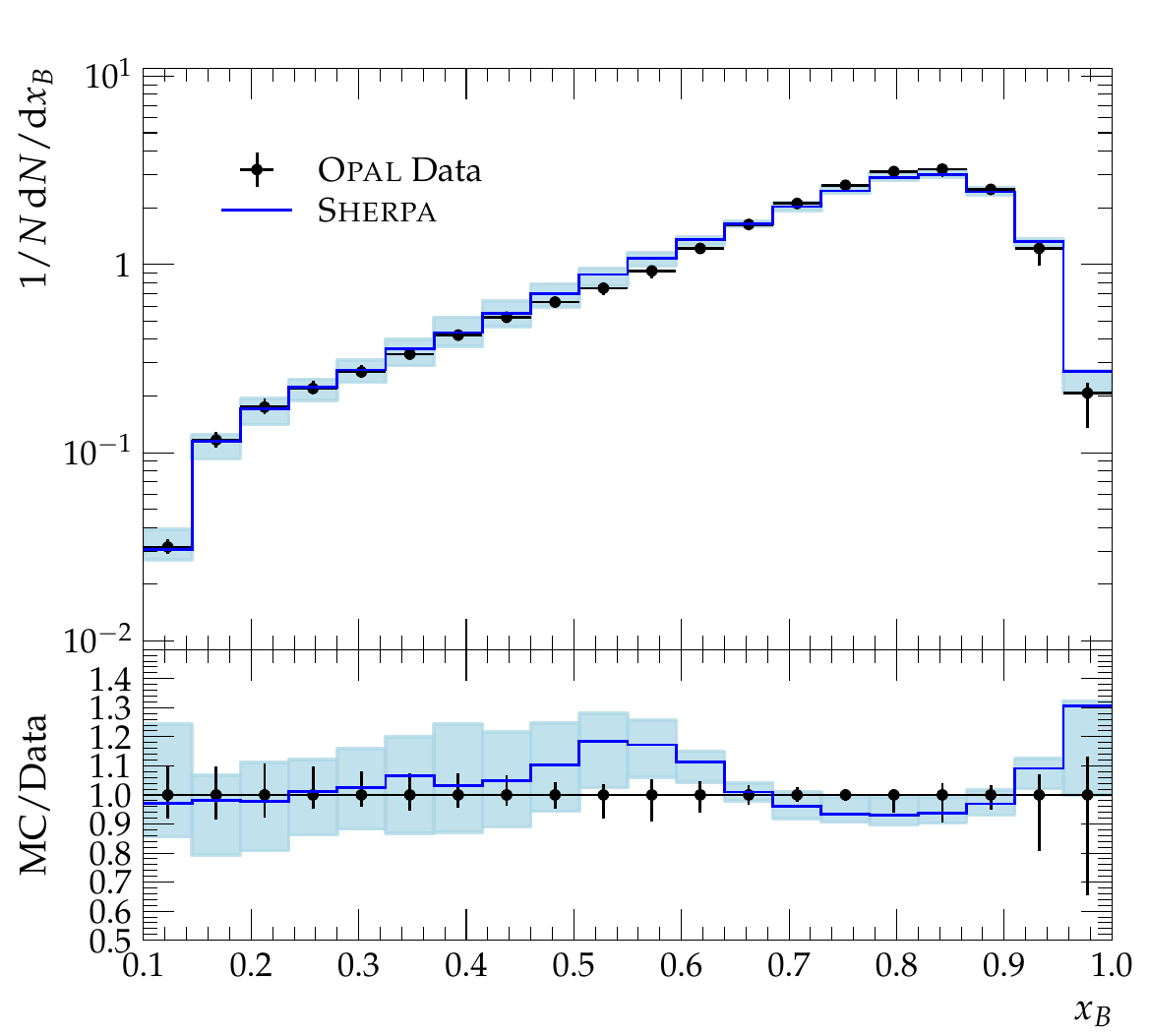}
  \caption{Example results for the \sherpa hadronisation tune including model parameter uncertainties in comparison to data from \LEP taken at $\sqrt{s}=91.2\,\text{GeV}$.
  Shown are the charged-particle multiplicity $n_\mathrm{ch}$ measured by \Aleph~\protect\cite{ALEPH:1991ldi} (left), thrust $T$ as measured by \Delphi~\protect\cite{DELPHI:1996sen} (center), and the $B$-hadron energy fraction $x_\mathrm{B}$ measured by \Opal~\protect\cite{OPAL:2002plk} (right).
   Each of the \sherpa predictions corresponds to $10^7$ events, ensuring that the statistical errors are negligible and the depicted uncertainties are dominated by the variations of non-perturbative model parameters.}
    \label{fig:tuning_plots}
\end{figure}

\section{Results}

Let us now turn to the statistical analysis of event shapes, measured as described in Sec.~\ref{sec:analysis}.
We add the histograms corresponding to our analysis to the \Contur framework \cite{Butterworth:2016sqg}, and use its statistical analysis modules to compute confidence levels for the exclusion of different points in the $(\mu_{gg},\mu_{q\bar{q}})$ plane.

\begin{figure}
    \includegraphics[width=.33\textwidth]{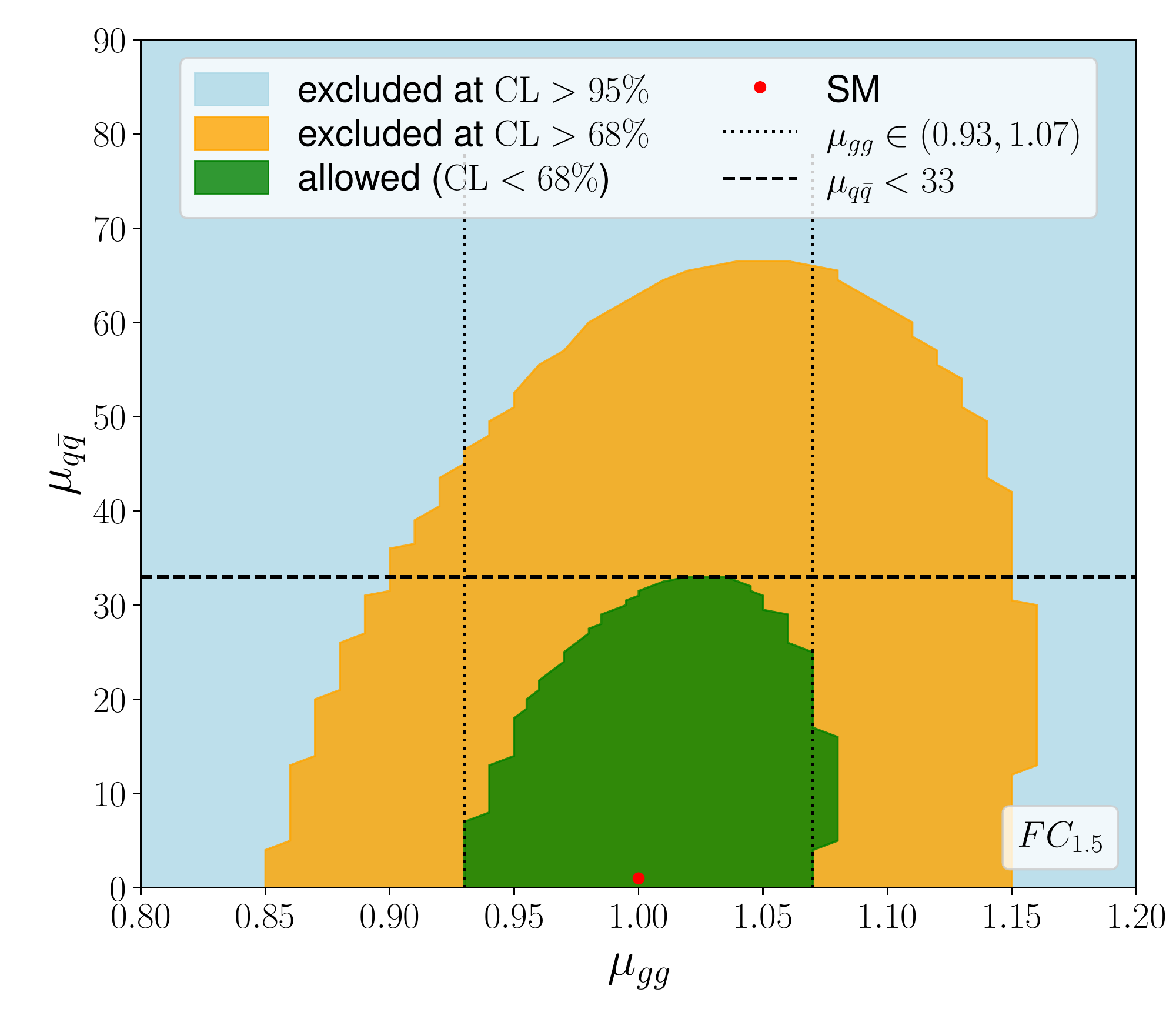}
    \includegraphics[width=.33\textwidth]{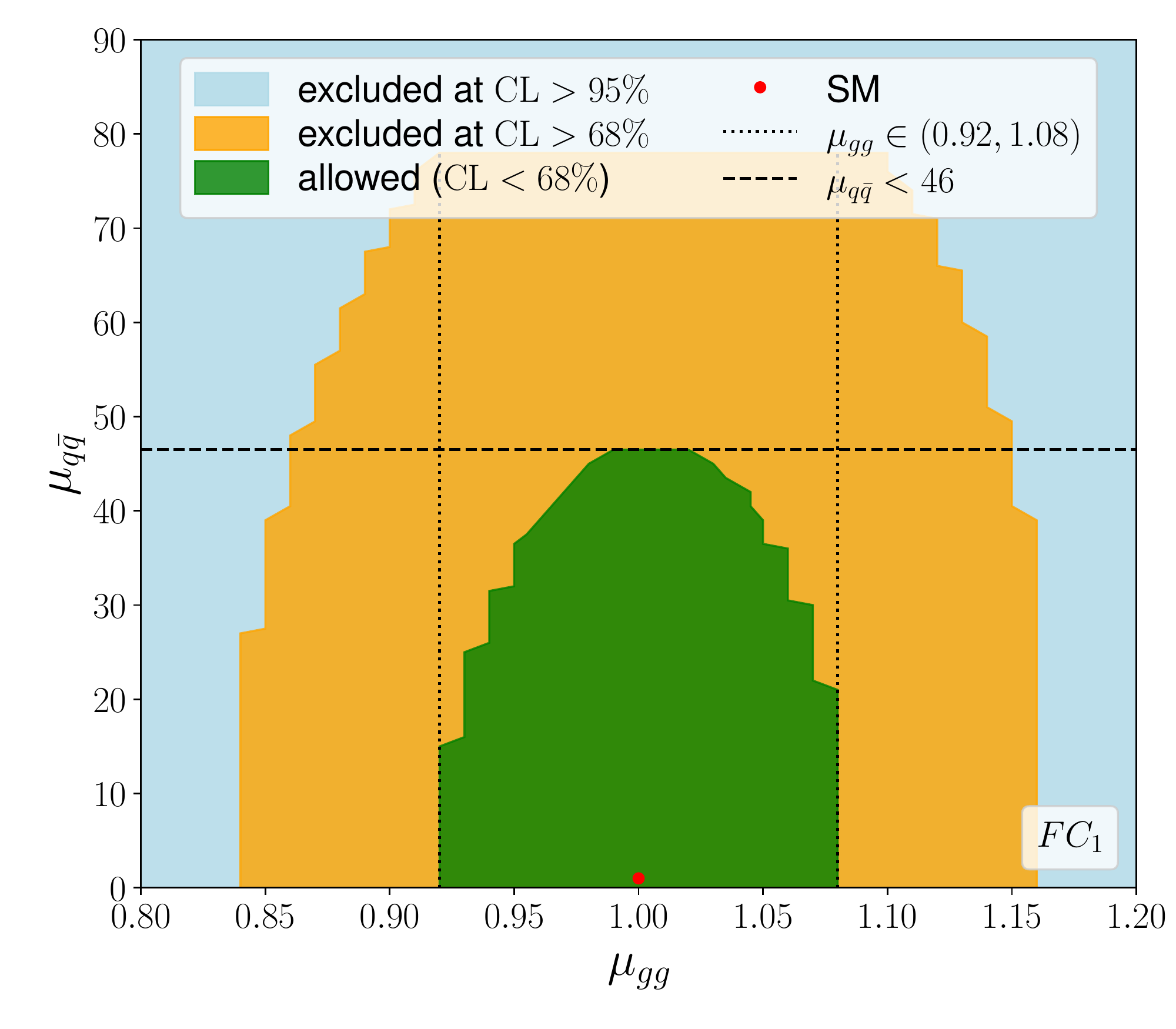}
    \includegraphics[width=.33\textwidth]{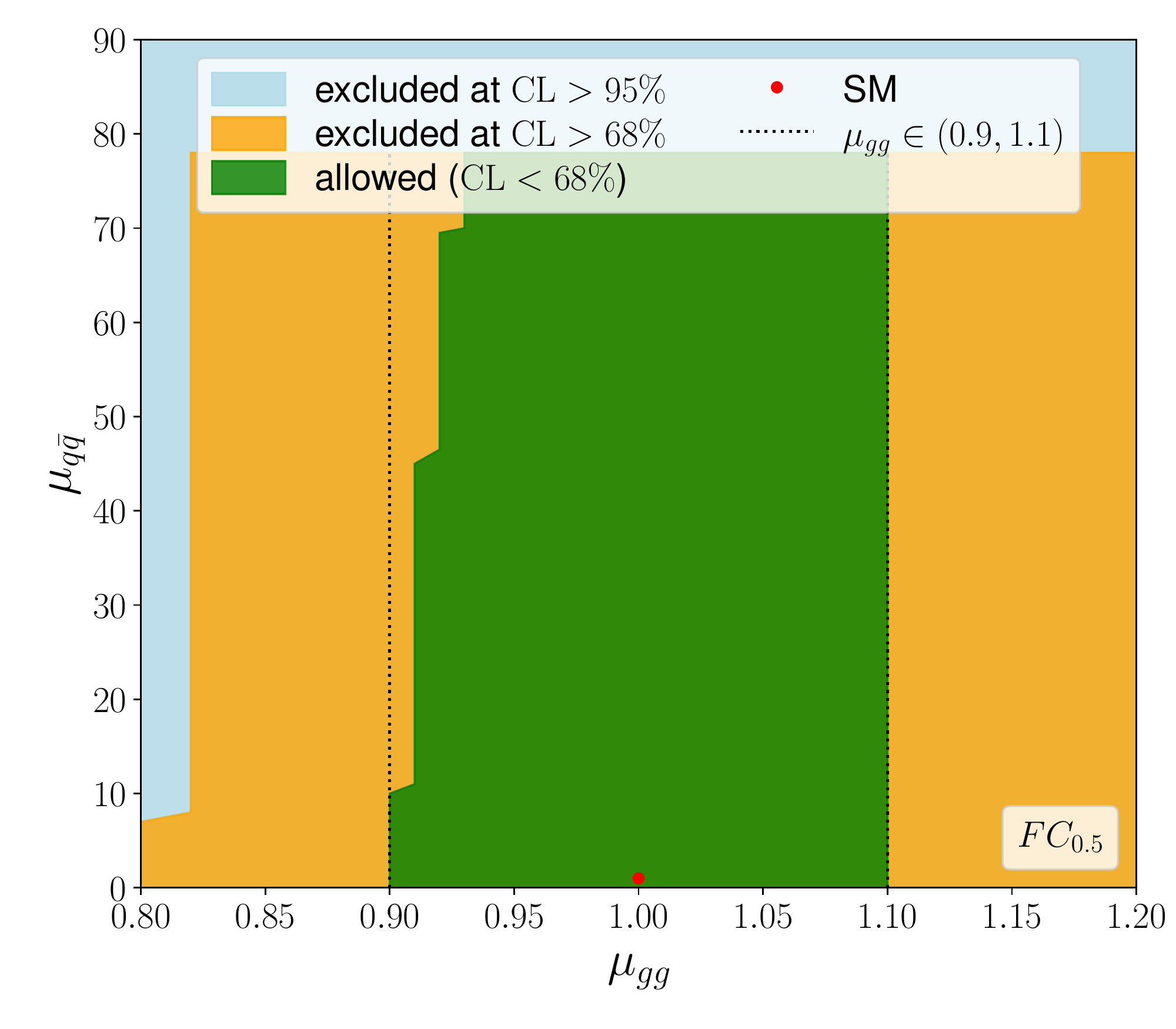}
    \caption{Exclusion limits based on fractional energy correlations (from left to right) $FC_{1.5}, FC_{1}, FC_{0.5}$.}\label{fig:cl_fcx}
\end{figure}
The two dimensional exclusion limits for $\mu_{gg}$ versus $\mu_{q\bar{q}}$ based on the three fractional energy correlations $FC_{1.5}$, $FC_{1}$ and $FC_{0.5}$ are shown in Fig.~\ref{fig:cl_fcx}, respectively.
As already indicated in the introduction, QCD radiation tends to mainly populate the soft and collinear regions of phase space, where the dead-cone effect associated to the finite and relatively large masses of the $c$ and $b$ quarks most visibly manifest themselves, and where differences due to different colour charges (the $C_F$ of quarks versus the $C_A$ of the gluons) lead to directly observable differences in the numbers of particles emitted. Accordingly, the Les-Houches angularity $FC_{1.5}$ tends to be the most sensitive observable, since it gives the largest weight to collinear emissions. Nevertheless, all three choices $x=1.5,1,0.5$ are able to limit $\mu_{gg}$ to be within $1\pm 0.10$ based on a $68\%$ confidence limit. However, based on $FC_{1.5}$ one should be able to set a stronger limit on $\mu_{gg}$ to be within $1\pm 0.07$ and additionally limit $\mu_{q\bar{q}} < 33$, while we can only exclude $\mu_{q\bar{q}}$ values larger than $45$ based on $FC_1$. Finally, $FC_{0.5}$ appears to not be sensitive to $\mu_{q\bar{q}}$ within the range we consider.

\begin{figure}
    \includegraphics[width=.33\textwidth]{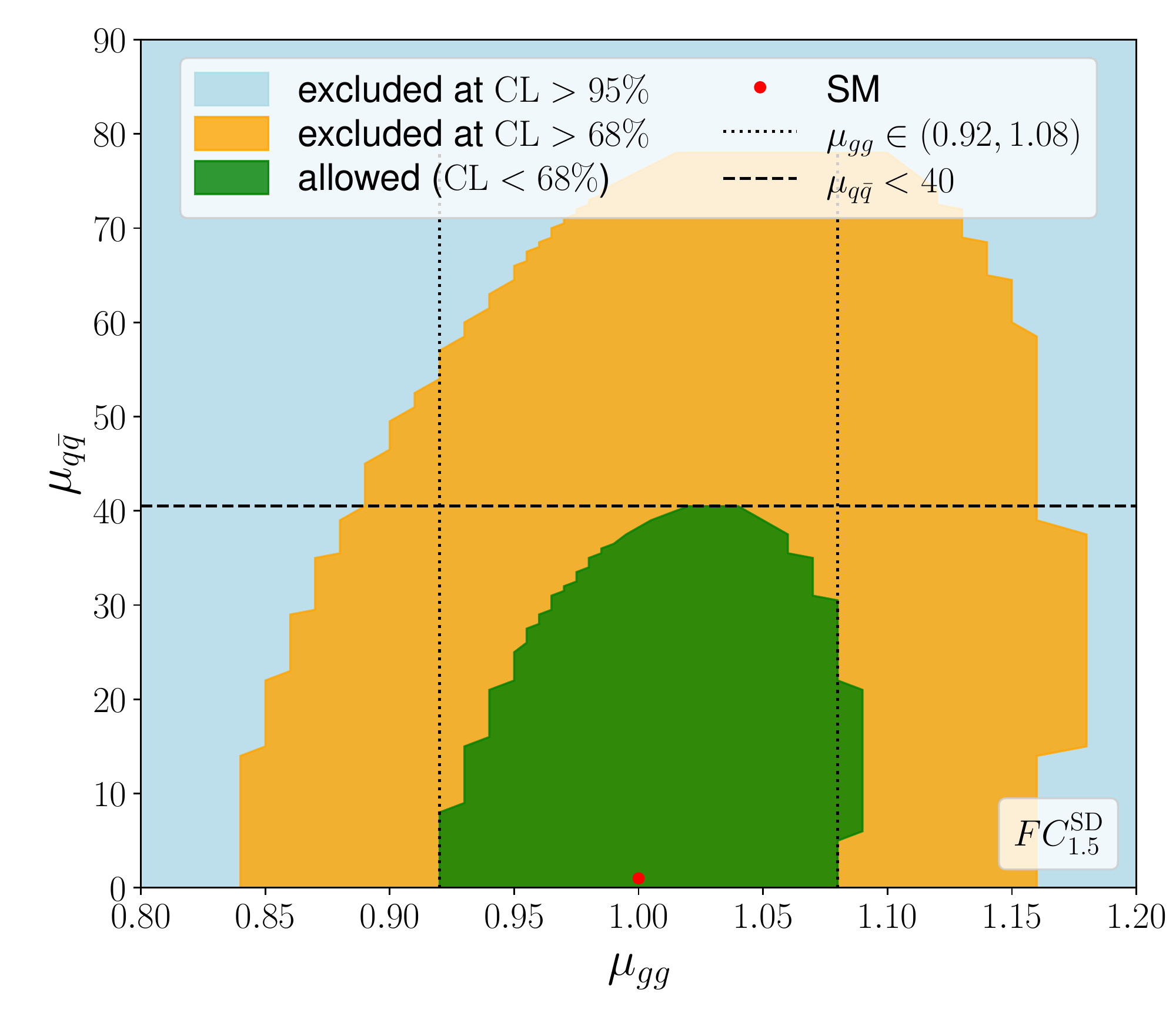}
    \includegraphics[width=.33\textwidth]{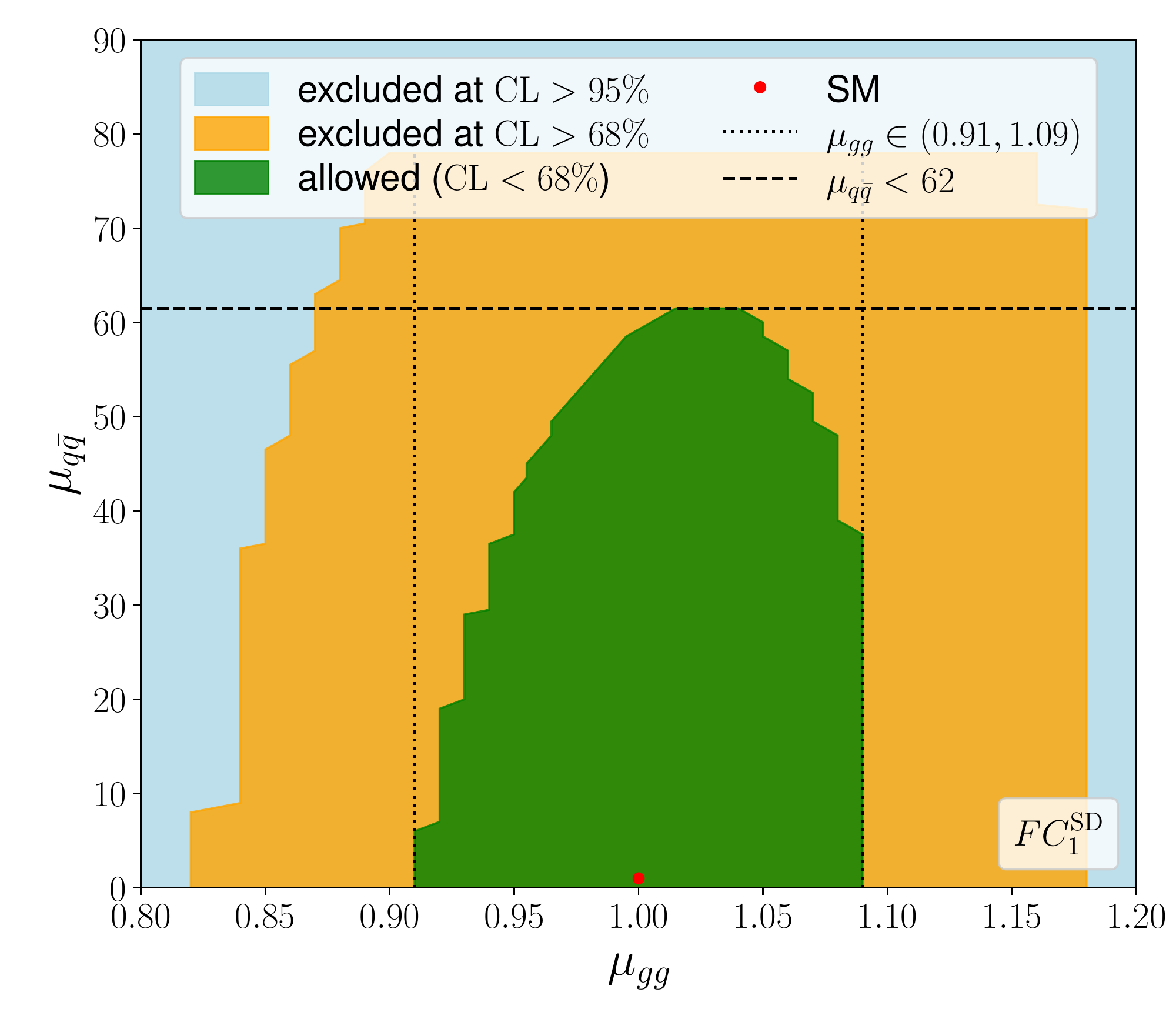}
    \includegraphics[width=.33\textwidth]{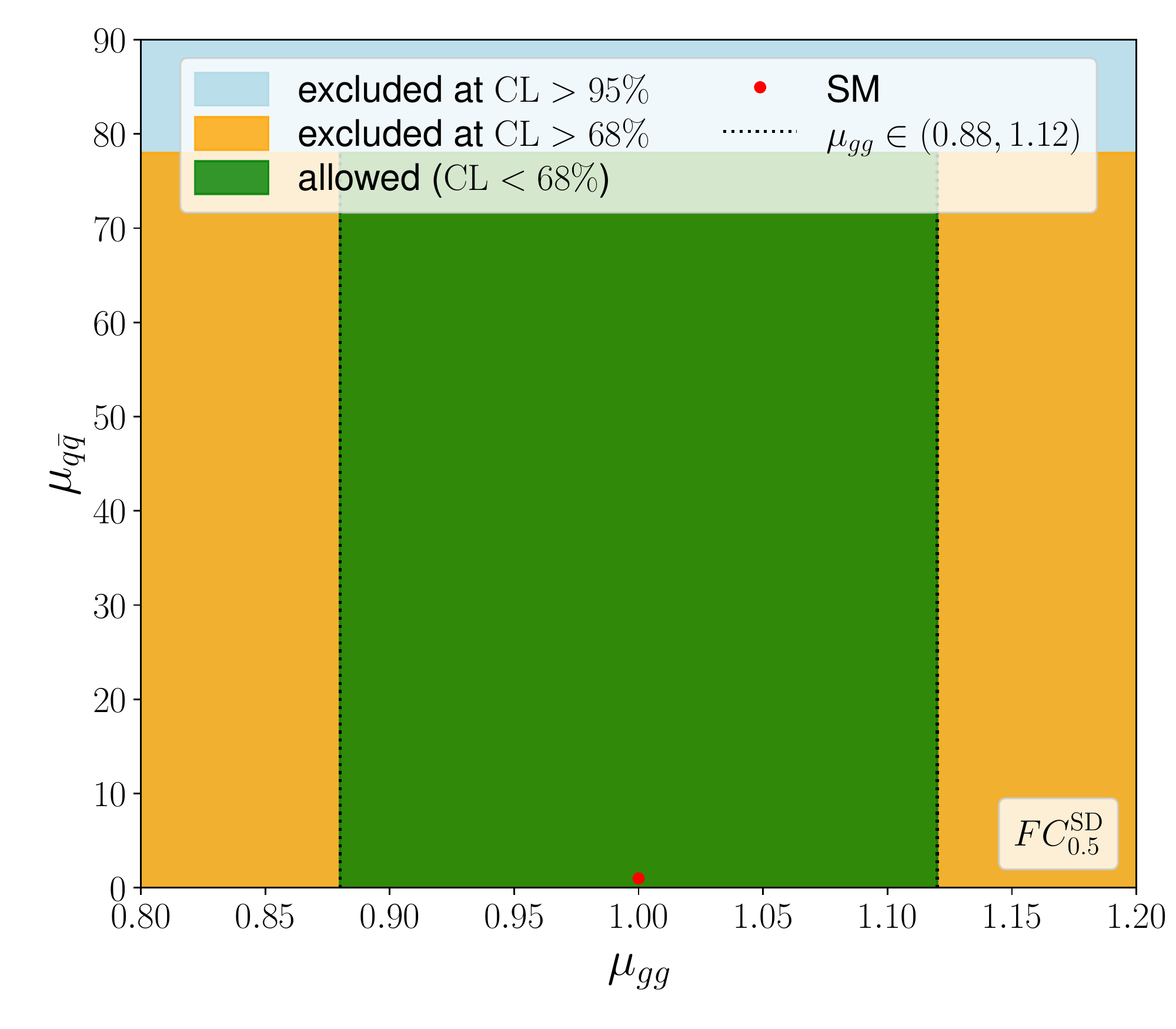}
    \caption{Exclusion limits based on soft-drop groomed fractional energy correlations (from left to right) $FC_{1.5}, FC_{1}, FC_{0.5}$.}\label{fig:cl_groom_fcx}
\end{figure}
Including soft-drop grooming of the hemispheres does not result in any significant improvements, as shown in Fig.~\ref{fig:cl_groom_fcx}, the equivalent of Fig.~\ref{fig:cl_fcx} but with grooming included.
In fact, the limits worsen slightly, which could to some extend have been anticipated, since there are competing effects at work.
Grooming will remove some information from the radiation pattern, but on the other hand has the potential to reduce the impact of hadronisation corrections and hence the associated systematic uncertainty.
Apparently, this reduction is not sufficient to compensate for the loss in information, at least with the grooming parameters we have considered here.
One could imagine that an optimisation of $z_\text{cut}$ and the inclusion of angular dependence in the soft-drop condition could lead to more competitive results. In addition it is certainly worth stressing that the combination of probable future refinements of the hadronisation models and the drastically increased data set of a potential FCC-ee ($10^{12}$ events vs $10^7$ at LEP-I) will most likely significantly reduce the uncertainty related to the modelling of the parton-to-hadron transition. To illustrate this, we present in Appendix~\ref{app:zeroNPuncertainty} Fig.~\ref{fig:no-NP-plots} 
selected exclusion-limit plots for the scenario of negligible non-perturbative uncertainties. As anticipated, the limits improve, resulting in 
$\mu_{gg}=1\pm0.05$ and $\mu_{q\bar{q}}<25$ for plain $FC_{1.5}$, and $\mu_{gg}=1\pm0.06$ and $\mu_{q\bar{q}}<28$ for its soft-drop groomed variant. 

\begin{figure}
    \includegraphics[width=.33\textwidth]{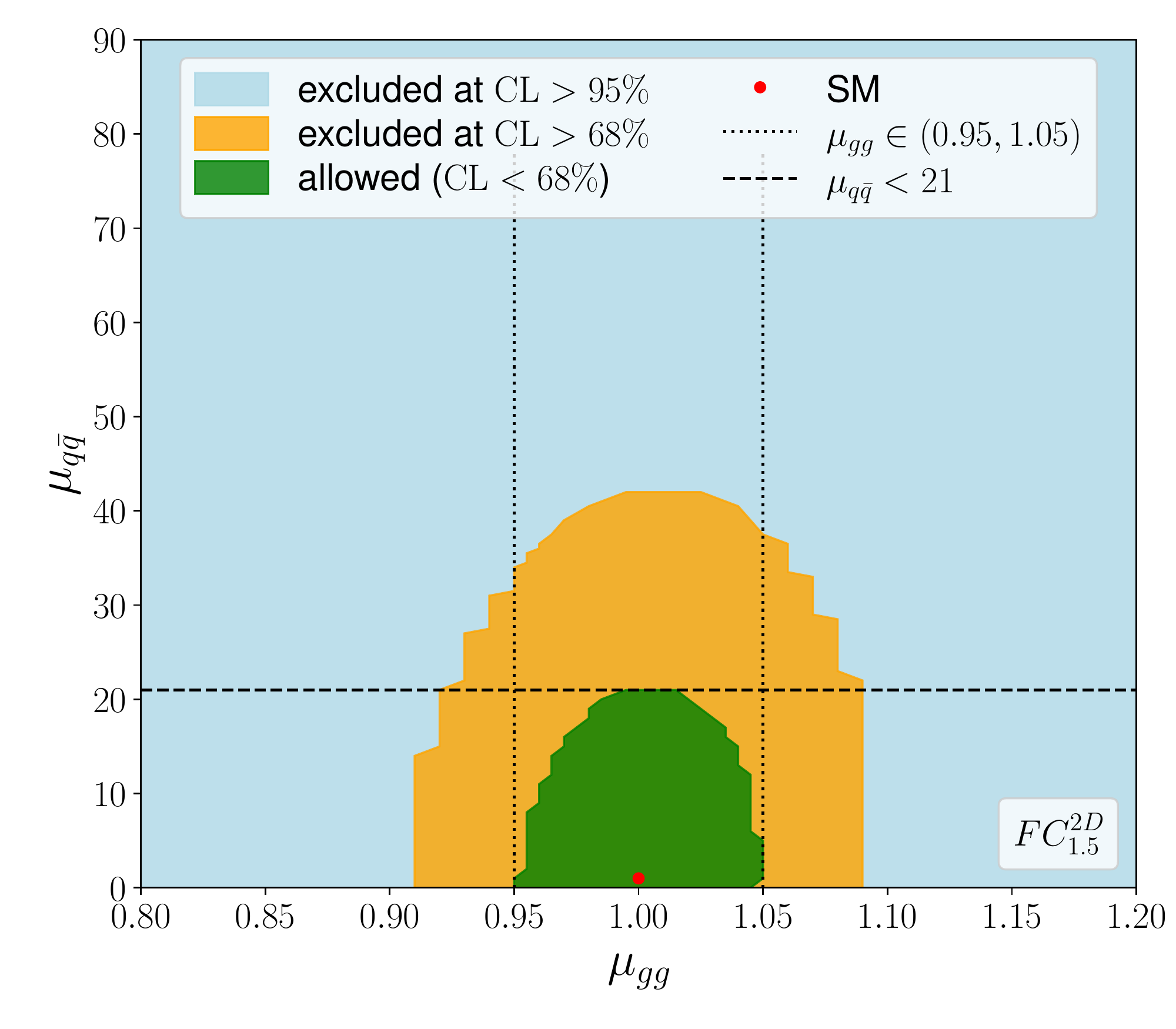}
    \includegraphics[width=.33\textwidth]{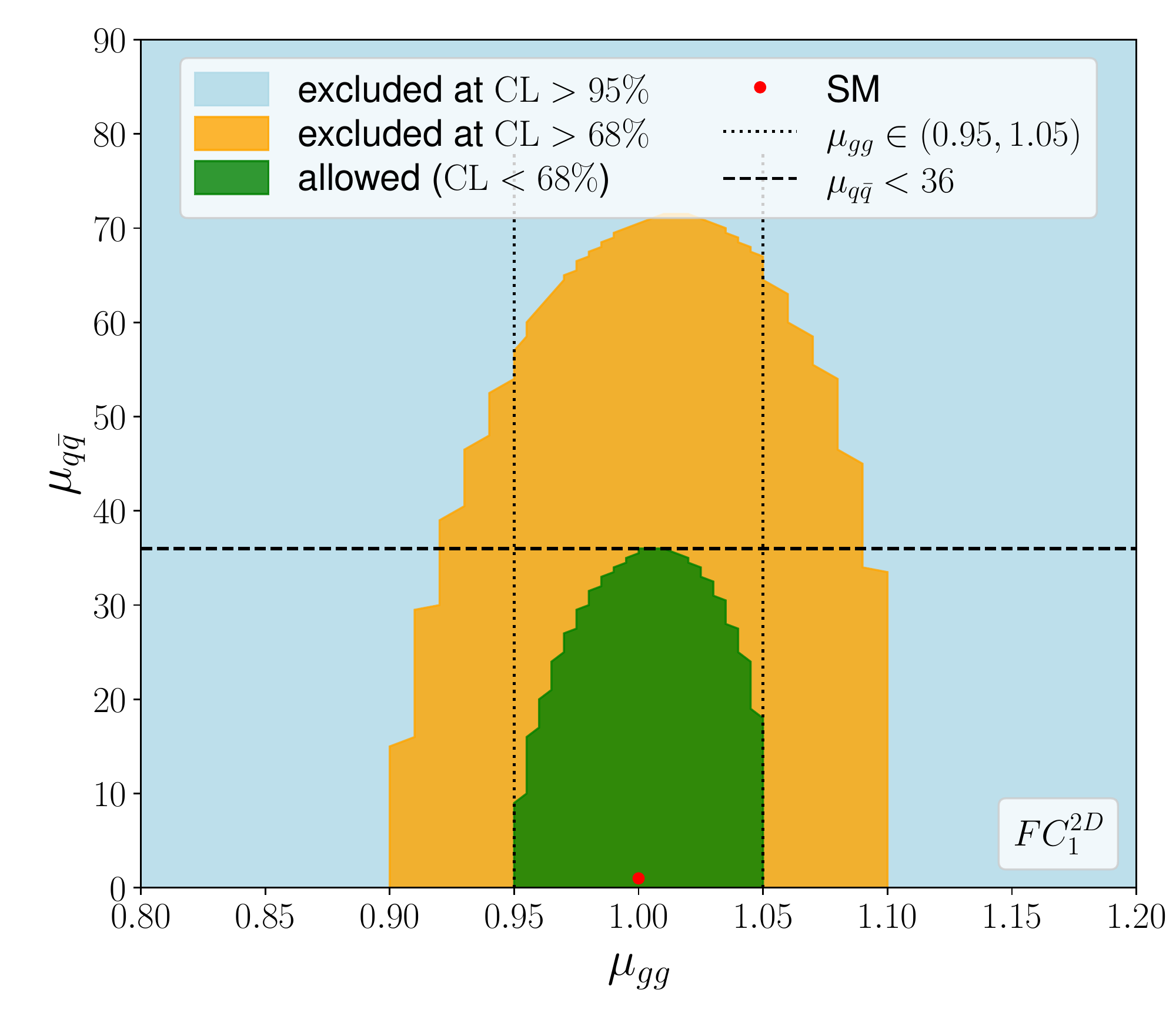}
    \includegraphics[width=.33\textwidth]{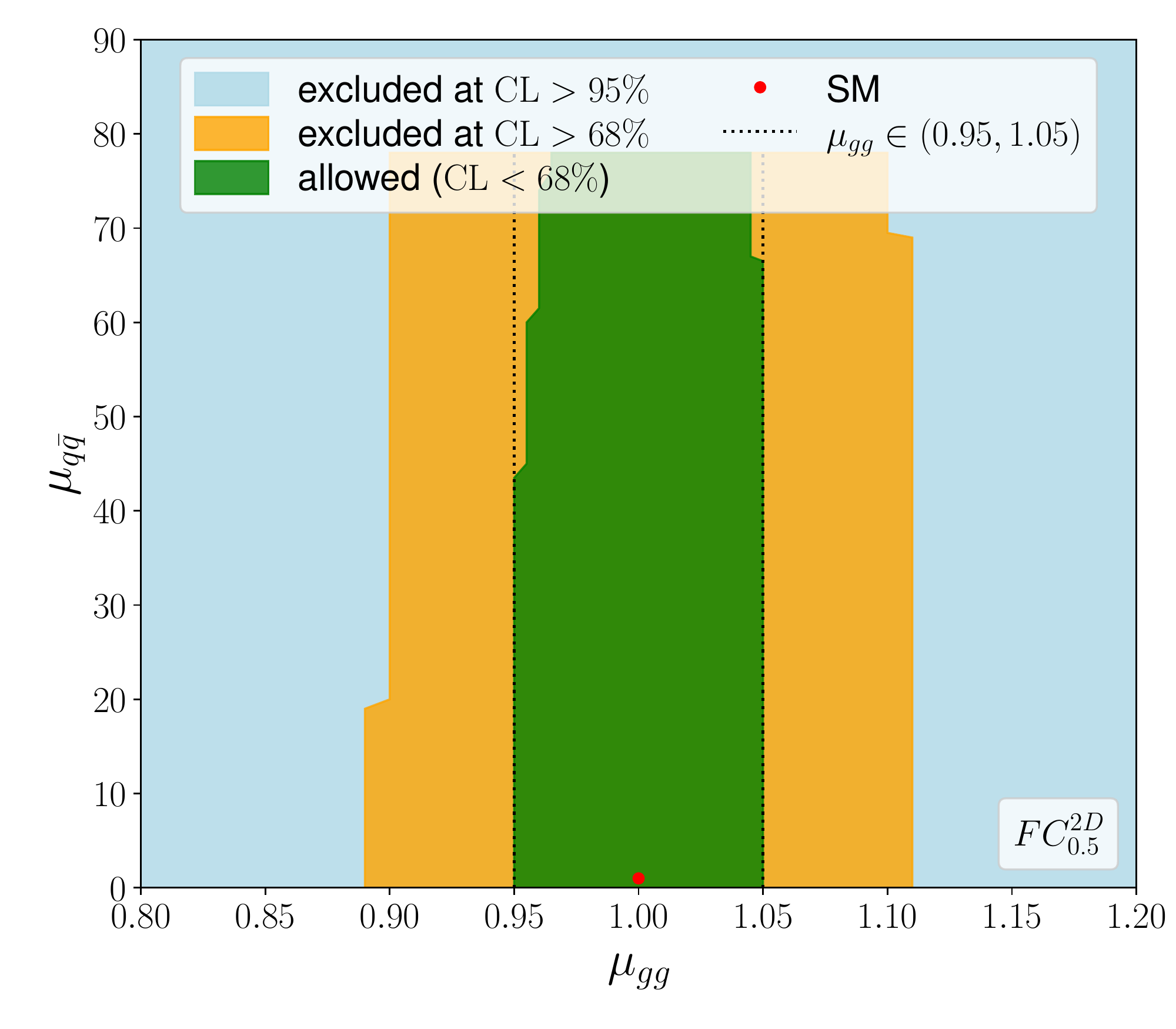}
    \caption{Exclusion limits based on fractional energy correlations (from left to right) $FC_{1.5}, FC_{1}, FC_{0.5}$, measured individually on the two hemispheres.}\label{fig:cl_2d_fcx}
\end{figure}
One of the major differences of our procedures so far, compared to traditional tagging methods, is that we effectively tag any event as a whole.
When individually tagging jets, or hemispheres for that matter, one would want to include a requirement that both tags are compatible with the desired final state.
To mimic this, we consider a measurement of the fractional energy correlations but on each hemisphere separately. We then derive exclusion limits based on the corresponding two-dimensional histograms.
While we hope to expose additional information in this way, it should be clear that this is a more involved observable definition.
In particular, joint resummed calculations of several observables are far less advanced than what would be available for the distributions considered above.
The resulting confidence levels can be found in Fig.~\ref{fig:cl_2d_fcx}. They are somewhat improved compared to the baseline in Fig.~\ref{fig:cl_fcx}. In particular, for $FC_{1.5}$ we obtain $\mu_{gg}=1\pm 0.05$ and $\mu_{q\bar{q}}<21$.
This suggests that combining individual results from the two hemispheres into one measurement reduces the impact of systematic uncertainties due to non-perturbative corrections, {\it i.e.}\ hadronization.  
This is further highlighted in the appendix, Fig.~\ref{fig:no-NP-plots} where we elucidate the impact of vanishing hadronization uncertainties due to the improved understanding and tuning of the corresponding models.

\begin{figure}
    \includegraphics[width=.33\textwidth]{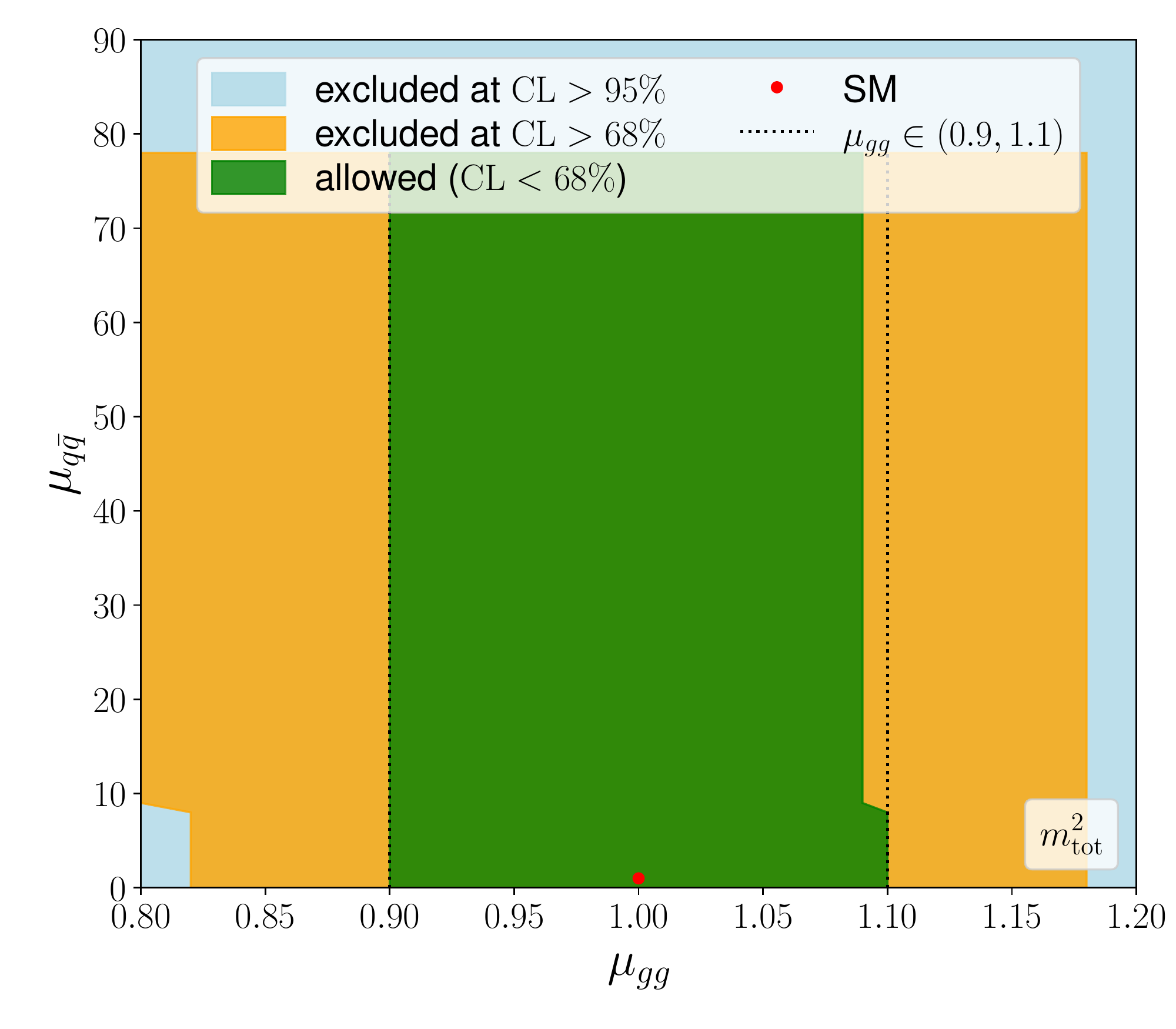}
    \includegraphics[width=.33\textwidth]{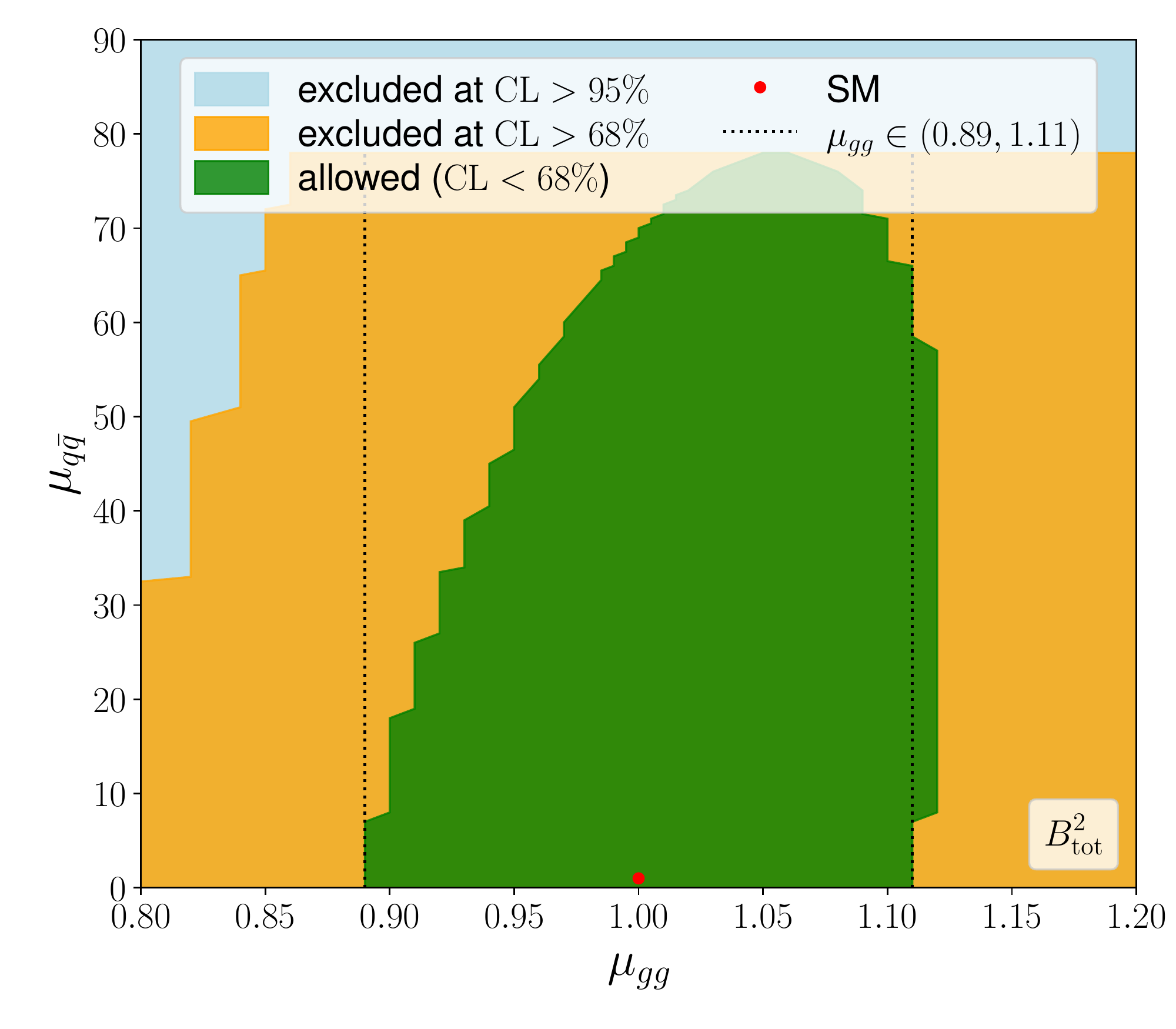}
    \includegraphics[width=.33\textwidth]{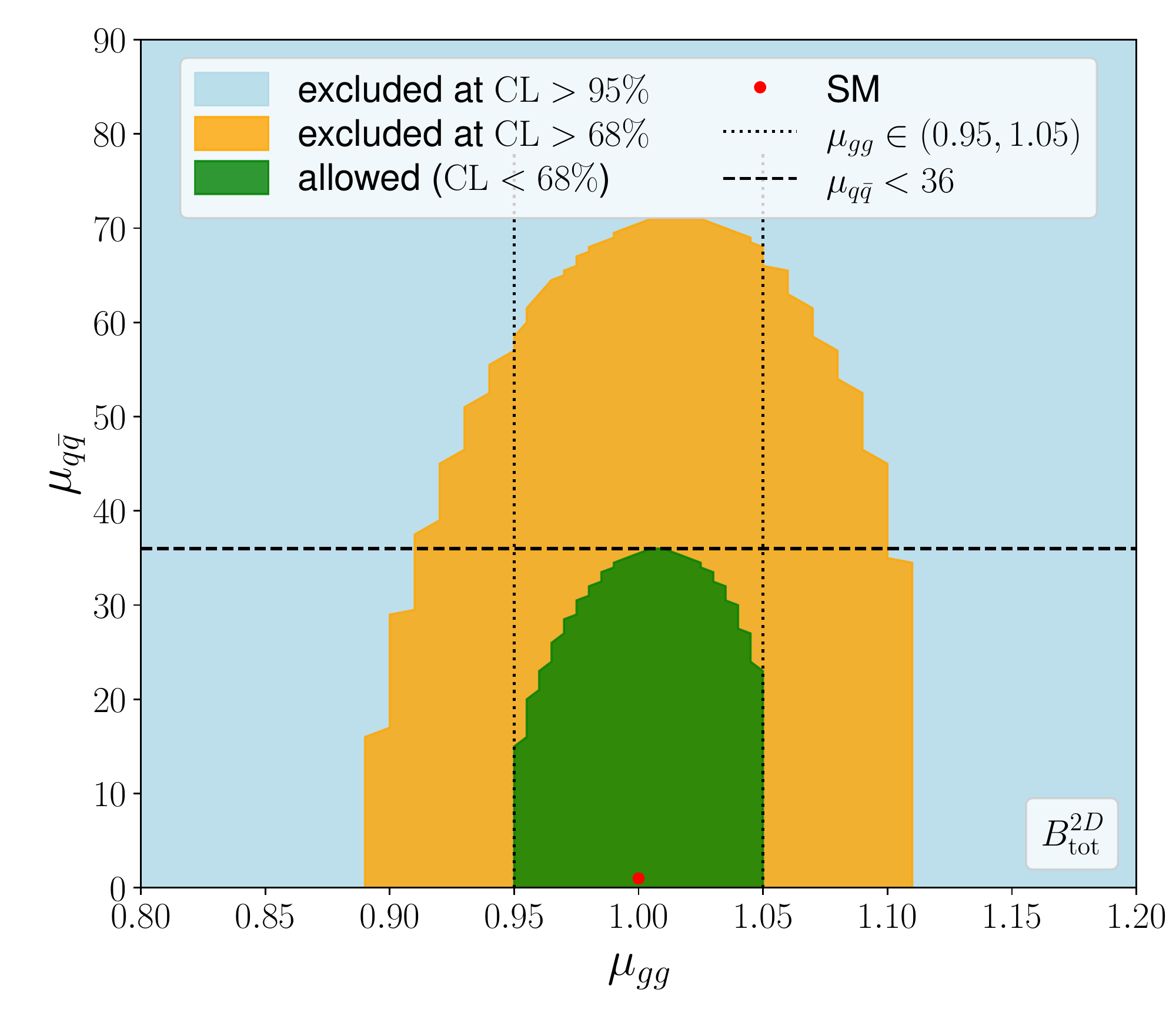}
    \caption{Exclusion limits based on (from left to right) the sum of the hemisphere masses, the total broadening, and the 2-dimensional distribution of broadenings in the two hemispheres.}\label{fig:cl_mass_broad}
\end{figure}

Lastly, we analyse some examples of more traditional event shapes that were measured by the \LEP experiments and enter our tuning. We focus on the sum of the hemisphere masses and the total broadening.
The hemisphere masses would scale similar to $FC_{0}$, which we do not show, in the infrared limit, while total broadening is similar to $FC_1$.
Correspondingly, the limits derived from the masses, shown in the leftmost subplot of Fig.~\ref{fig:cl_mass_broad}, is even weaker than that obtained from $FC_{0.5}$, consistent with the decline depending on $x$.
In the middle of Fig.~\ref{fig:cl_mass_broad} we show the same plot but for total broadening, where we find a similar behaviour to $FC_1$, as expected.
Finally, in the rightmost plot we show the limits obtained from considering the two-dimensional distribution of the broadening of both hemispheres.
Again, we observe slightly improved limits compared to the one-dimensional distribution, in line with what was observed for $FC_1$.

\section{Conclusions}

In this article we have presented a proposal for the extraction of hadronic branching ratios of the Higgs boson at a future lepton collider operated as a Higgs factory. 
To separate couplings of the Higgs boson to light degrees of freedom, \emph{i.e.}\ those to gluons and light quarks from the ones to massive charm and bottom quarks, our ansatz does not require the reconstruction of displaced vertices from weak decays of heavy flavour hadrons. 
Instead, we propose to instrument event-shape observables that are sensitive to the differences in the radiation patterns of light/heavy quarks and gluons. 
In particular, we have focused on fractional energy correlations with and without soft-drop grooming of the two event hemispheres with respect to the thrust axis. 
Based on dedicated simulations with the \sherpa event generator framework we showed that stringent limits on the deviation of the Higgs-boson branching ratios into light-quarks, gluons and bottom-quarks can be obtained.
Assuming the full projected luminosity of $5\,\text{ab}^{-1}$ for the FCC-ee collider at $\sqrt{s}=240\,\text{GeV}$, we were able to derive limits as tight as
\begin{equation*}
\mu_{gg}=1\pm 0.05\,,\quad\mu_{q\bar{q}}<21\,,\quad\text{while Eq.~\eqref{eq:mu_const} has to hold}.
\end{equation*}
The best sensitivity we obtained for the fractional energy correlation $FC_{1.5}$ that is geared to explore the collinear region of the hard shower initiator. 
To address the systematic uncertainty from the non-perturbative fragmentation of partons to hadrons we performed a dedicated tune to event-shape and jet observables measured by the \LEP experiments, thereby deriving, for the first time, a whole family of replica tunes that allow us to estimate the residual model-parameter uncertainties of \sherpa's cluster fragmentation.

The obtained results motivate further studies and refinements. 
In our study, we deliberately ignored any information on potential secondary vertices that could in principle be used for a better separation of the heavy-flavour contributions. 
Furthermore, additional selection cuts could help to further reduce the contribution from Higgs-boson decays to vector bosons that subsequently decay hadronically, populating the region of rather large event-shape values, thereby diluting the gluon signal. 
Similarly, the parameters of the grooming algorithm can certainly be further optimised to better balance the reduction of non-perturbative corrections against the imprints of the QCD radiation pattern in the hadronic final states.

Concerning the theoretical predictions, for the considered event-shape variables all-orders analytical predictions could be derived, for example at next-leading-logarithmic accuracy based on the \Caesar formalism~\cite{Banfi:2004yd,Gerwick:2014gya,Marzani:2019evv,Baberuxki:2019ifp}, or at next-to-next-to-leading-logarithmic level through \ARES~\cite{Banfi:2014sua,Banfi:2016zlc}. 
Alternatively, accurate resummed predictions for event shapes have been obtained using effective field theory techniques, see for example \cite{Abbate:2010xh, Hoang:2014wka}, which have also been applied to jet observables closely related to the energy correlations used here \cite{Ellis:2010rwa, Larkoski:2013eya, Larkoski:2014uqa, Hornig:2016ahz, Kang:2018qra}.
Furthermore, progress has recently been made on including finite quark masses in resummed calculations~\cite{Fickinger:2016rfd, Gaggero:2022hmv, Ghira:2023bxr}. 
To further reduce non-perturbative uncertainties we envisage dedicated analyses of hadronic final states at the FCC-ee prior to attempts to measure Higgs-boson couplings that should enter the tuning and help to further constrain the model-parameter uncertainties.

\section*{Acknowledgements}
The work of MK and SS was supported by BMBF (contract 05H21MGCAB) and DFG (project 456104544, 510810461) and FK and DR were supported by the STFC IPPP grant (ST/T001011/1).
FK also acknowledges funding as Royal Society Wolfson Research fellow.

\appendix
\section{Exclusion limits assuming vanishing non-perturbative uncertainties}
\label{app:zeroNPuncertainty}

\begin{figure}[h!]
    \centering
    \includegraphics[width=0.32\textwidth]{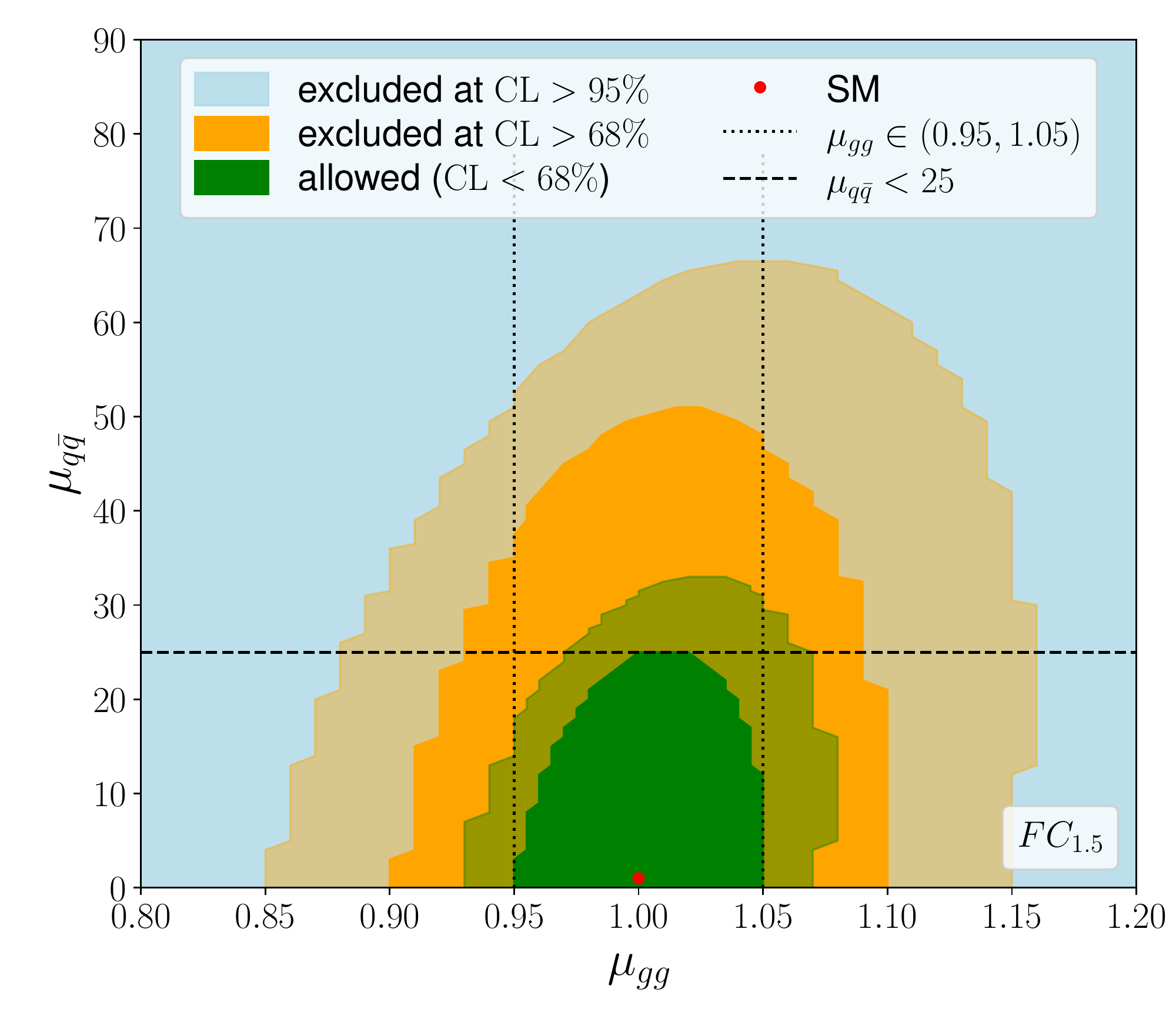}
    \includegraphics[width=0.32\textwidth]{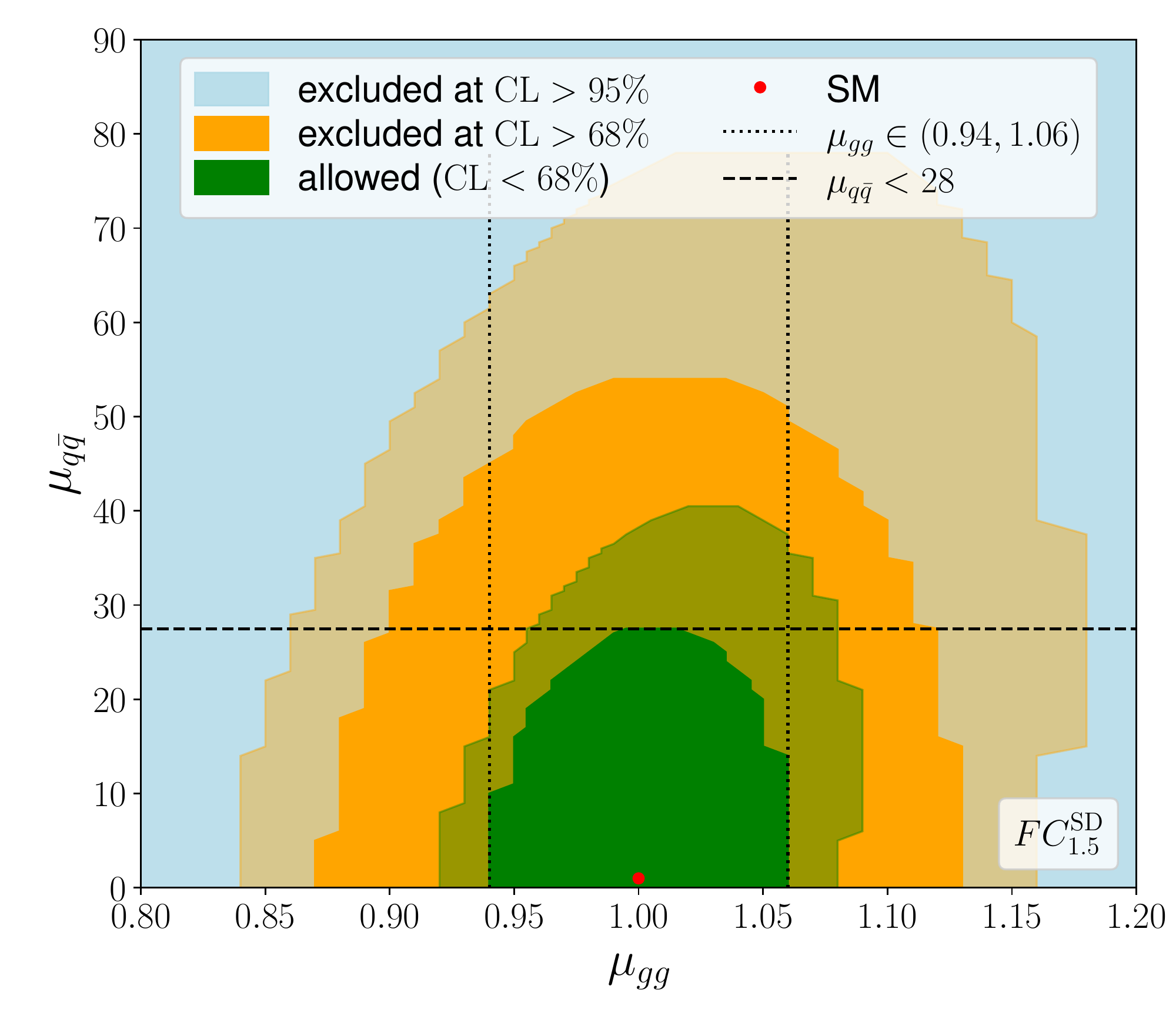}
    \includegraphics[width=0.32\textwidth]{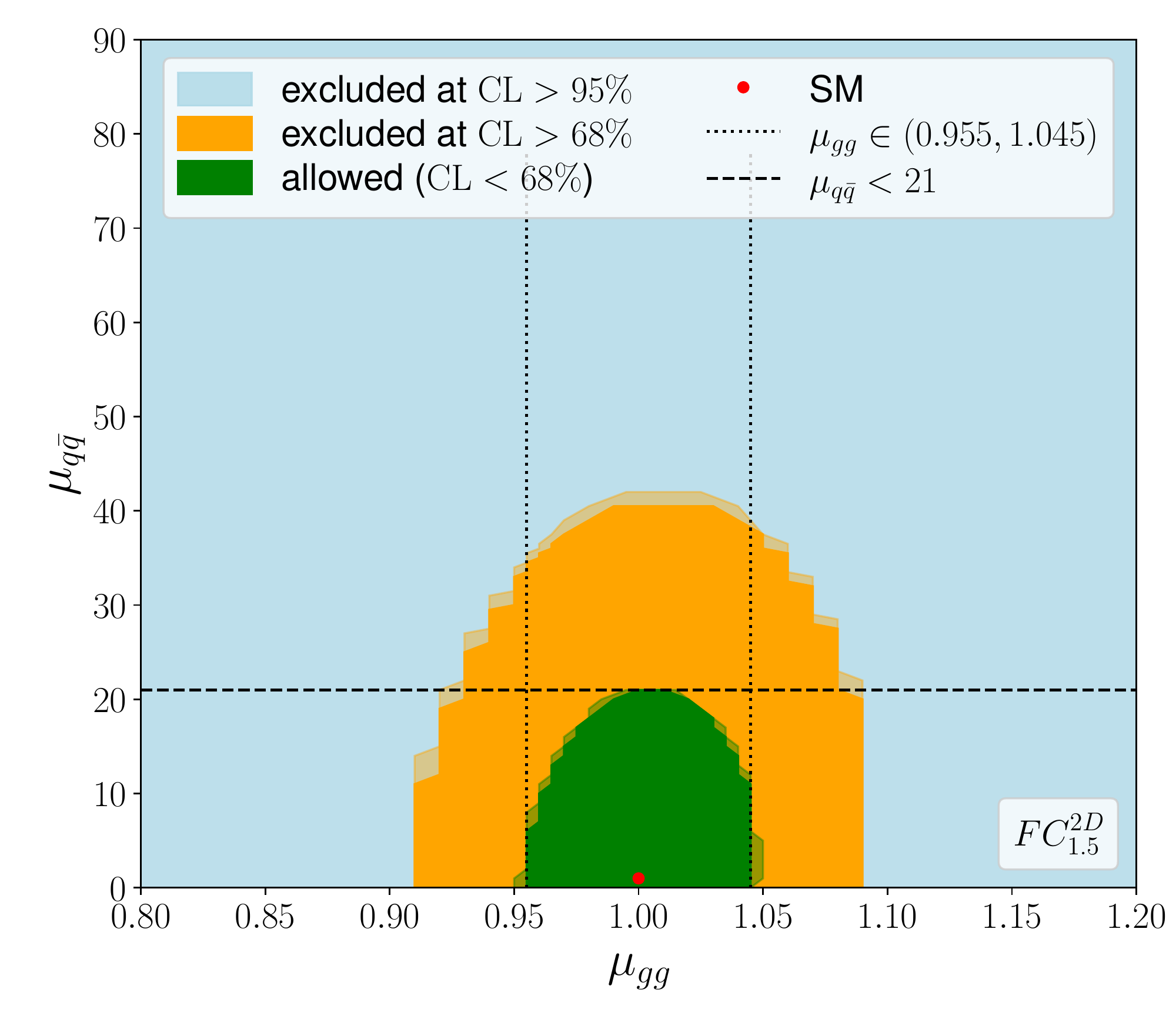}
    \caption{Examples of limits obtained from variants of the $FC_{1.5}$ observable when neglecting non-perturbative uncertainties, without and with soft-drop grooming (left, middle) and measured individually on the two event hemispheres (right). The transparent bounds are identical to the main text, while the solid ones are obtained when including statistical uncertainties only.}
    \label{fig:no-NP-plots}
\end{figure}

We highlight the effect of vanishing non-perturbative  uncertainties in Fig.~\ref{fig:no-NP-plots}, where we show exclusion plots for the three procedural variants of the $FC_{1.5}$ observable considered before. 

\section{Central tune parameters and uncertainty ranges}
\label{app:tuning_details}
The re-tuning of \sherpa's hadronisation model was based on \rivet analyses of \LEP1 measurements and observables detailed in Tab.~\ref{tab:tuning_analyses}.
We list the the initially considered parameter range, the central tune value and corresponding uncertainty intervals, determined from 50 replica tunes, in Tab.~\ref{tab:tuning_results}.
For the full list of \Ahadic model parameters and their physical interpretation we refer the reader to Ref.~\cite{Chahal:2022rid}.
\begin{table}[h!]
    \centering
    \begin{tabular}{l|c}
        \rivet analysis tag [reference] & observables \\\hline
        \texttt{ALEPH\_1991\_S2435284} \cite{ALEPH:1991ldi} & mean/total charged multiplicity\\
        \texttt{ALEPH\_1999\_S4193598} \cite{ALEPH:1999syy} & scaled energy of $D^{*\pm}$ mesons\\
        \texttt{OPAL\_2003\_I599181} \protect\cite{OPAL:2002plk} & $b$-quark fragmentation function\\
        \texttt{ALEPH\_2004\_S5765862} \cite{ALEPH:2003obs} & jet-mass difference, aplanarity, oblateness, C-parameter\\
        \texttt{DELPHI\_1996\_S3430090} \cite{DELPHI:1996sen} & thrust (major, minor), $p_\perp$ w.r.t. thrust axis, scaled momentum\\
        \texttt{JADE\_OPAL\_2000\_S4300807} \cite{JADE:1999zar} & Durham algorithm differential jet rate $y_{2}$\\
        \texttt{PDG\_HADRON\_MULTIPLICITIES} \cite{ParticleDataGroup:2008zun} & multiplicity of $\pi^+, \pi^0, K^+, K^0, \eta,\eta',p,\Lambda$ hadrons\\
    \end{tabular}
    \caption{List of \rivet analyses and corresponding observables used for the tuning.}
    \label{tab:tuning_analyses}
\end{table}

\begin{table}[h!]
    \centering
    \begin{tabular}{c|c|c|c}
        Parameter tag                   & tuning range & central tune & uncertainty variation \\\hline
        \texttt{KT\_0}                  & $[0.5,1.5   ]$ & 1.21  &  $[1.15, 1.35]   $\\
        \texttt{ALPHA\_G}               & $[-1.0, 2.0 ]$ & 0.96  &  $[0.50, 1.01]   $\\
        \texttt{ALPHA\_L}               & $[-1.0,5.0  ]$ & 3.92 &  $[3.88, 4.27]   $\\
        \texttt{BETA\_L}                & $[0.0, 0.5  ]$ & 0.18  &  $[0.05, 0.20]   $\\
        \texttt{GAMMA\_L}               & $[0.1,1.0   ]$ & 0.47  &  $[0.34, 0.65]   $\\
        \texttt{ALPHA\_D}               & $[-1.0,5.0  ]$ & 3.41  &  $[3.10, 4.17]   $\\
        \texttt{BETA\_D}                & $[0.0,1.0   ]$ & 0.72 &  $[0.58, 0.83]   $\\
        \texttt{GAMMA\_D}               & $[0.0,1.0   ]$ & 0.77  &  $[0.51, 0.82]   $\\
        \texttt{ALPHA\_H}               & $[-1.0,5.0  ]$ & -0.60 &  $[-0.70, -0.15] $\\
        \texttt{BETA\_H}                & $[0.1, 2.0  ]$ & 1.84  &  $[1.53, 1.90]   $\\
        \texttt{GAMMA\_H}               & $[0.0,0.5   ]$ & 0.024 &  $[0.021, 0.049] $\\
        \texttt{STRANGE\_FRACTION}      & $[0.0,1.0   ]$ & 0.46  &  $[0.43, 0.48]   $\\
        \texttt{BARYON\_FRACTION}       & $[0.0,1.0   ]$ & 0.17  &  $[0.15, 0.18]   $\\
        \texttt{P\_QS\_by\_P\_QQ\_norm} & $[0.0,1.0   ]$ & 0.57  &  $[0.54, 0.77]   $\\
        \texttt{P\_SS\_by\_P\_QQ\_norm} & $[0.0,0.1   ]$ & 0.056 &  $[0.01, 0.08]   $\\
        \texttt{P\_QQ1\_by\_P\_QQ0}     & $[0.0,2.0   ]$ & 0.60  &  $[0.53, 0.71]   $\\
    \end{tabular}
    \caption{List of \Ahadic model parameters considered in the tuning.
    Quoted are the initial parameter
    interval, the central-tune value, and the parameter uncertainties extracted from 50 replica tunes. }
    \label{tab:tuning_results}
\end{table}
\clearpage
\bibliographystyle{SciPost_bibstyle}
\bibliography{refs}

\begin{thebibliography}{10}
\providecommand{\url}[1]{\texttt{#1}}
\providecommand{\urlprefix}{URL }
\expandafter\ifx\csname urlstyle\endcsname\relax
  \providecommand{\doi}[1]{doi:\discretionary{}{}{}#1}\else
  \providecommand{\doi}{doi:\discretionary{}{}{}\begingroup
  \urlstyle{rm}\Url}\fi
\providecommand{\eprint}[2][]{\url{#2}}

\bibitem{FCC:2018byv}
A.~Abada \emph{et~al.},
\newblock \emph{{FCC Physics Opportunities}: {Future Circular Collider
  Conceptual Design Report Volume 1}},
\newblock Eur. Phys. J. C \textbf{79}(6), 474 (2019),
\newblock \doi{10.1140/epjc/s10052-019-6904-3}.

\bibitem{FCC:2018evy}
A.~Abada \emph{et~al.},
\newblock \emph{{FCC-ee: The Lepton Collider}: {Future Circular Collider
  Conceptual Design Report Volume 2}},
\newblock Eur. Phys. J. ST \textbf{228}(2), 261 (2019),
\newblock \doi{10.1140/epjst/e2019-900045-4}.

\bibitem{Weinberg:1979sa}
S.~Weinberg,
\newblock \emph{{Baryon and Lepton Nonconserving Processes}},
\newblock Phys. Rev. Lett. \textbf{43}, 1566 (1979),
\newblock \doi{10.1103/PhysRevLett.43.1566}.

\bibitem{Buchmuller:1985jz}
W.~Buchm{\"u}ller and D.~Wyler,
\newblock \emph{{Effective Lagrangian Analysis of New Interactions and Flavor
  Conservation}},
\newblock Nucl. Phys. B \textbf{268}, 621 (1986),
\newblock \doi{10.1016/0550-3213(86)90262-2}.

\bibitem{Grzadkowski:2010es}
B.~Grzadkowski, M.~Iskrzynski, M.~Misiak and J.~Rosiek,
\newblock \emph{{Dimension-Six Terms in the Standard Model Lagrangian}},
\newblock JHEP \textbf{10}, 085 (2010),
\newblock \doi{10.1007/JHEP10(2010)085},
\newblock \eprint{1008.4884}.

\bibitem{Azzi:2012yn}
P.~Azzi, C.~Bernet, C.~Botta, P.~Janot, M.~Klute, P.~Lenzi, L.~Malgeri and
  M.~Zanetti,
\newblock \emph{{Prospective Studies for LEP3 with the CMS Detector}}  (2012),
\newblock \eprint{1208.1662}.

\bibitem{CMS:2018nsn}
A.~M. Sirunyan \emph{et~al.},
\newblock \emph{{Observation of Higgs boson decay to bottom quarks}},
\newblock Phys. Rev. Lett. \textbf{121}(12), 121801 (2018),
\newblock \doi{10.1103/PhysRevLett.121.121801},
\newblock \eprint{1808.08242}.

\bibitem{ATLAS:2020jwz}
G.~Aad \emph{et~al.},
\newblock \emph{{Measurement of the associated production of a Higgs boson
  decaying into $b$-quarks with a vector boson at high transverse momentum in
  $pp$ collisions at $\sqrt{s} = 13$ TeV with the ATLAS detector}},
\newblock Phys. Lett. B \textbf{816}, 136204 (2021),
\newblock \doi{10.1016/j.physletb.2021.136204},
\newblock \eprint{2008.02508}.

\bibitem{CMS:2019hve}
A.~M. Sirunyan \emph{et~al.},
\newblock \emph{{A search for the standard model Higgs boson decaying to charm
  quarks}},
\newblock JHEP \textbf{03}, 131 (2020),
\newblock \doi{10.1007/JHEP03(2020)131},
\newblock \eprint{1912.01662}.

\bibitem{ATLAS:2022ers}
G.~Aad \emph{et~al.},
\newblock \emph{{Direct constraint on the Higgs-charm coupling from a search
  for Higgs boson decays into charm quarks with the ATLAS detector}},
\newblock Eur. Phys. J. C \textbf{82}, 717 (2022),
\newblock \doi{10.1140/epjc/s10052-022-10588-3},
\newblock \eprint{2201.11428}.

\bibitem{deBlas:2019rxi}
J.~de~Blas \emph{et~al.},
\newblock \emph{{Higgs Boson Studies at Future Particle Colliders}},
\newblock JHEP \textbf{01}, 139 (2020),
\newblock \doi{10.1007/JHEP01(2020)139},
\newblock \eprint{1905.03764}.

\bibitem{Walker:2022yml}
J.~Walker and F.~Krauss,
\newblock \emph{{Constraining the Charm-Yukawa coupling at the Large Hadron
  Collider}},
\newblock Phys. Lett. B \textbf{832}, 137255 (2022),
\newblock \doi{10.1016/j.physletb.2022.137255},
\newblock \eprint{2202.13937}.

\bibitem{Amoroso:2020lgh}
S.~Amoroso \emph{et~al.},
\newblock \emph{{Les Houches 2019: Physics at TeV Colliders: Standard Model
  Working Group Report}},
\newblock In \emph{{11th Les Houches Workshop on Physics at TeV Colliders}:
  {PhysTeV Les Houches}} (2020), \eprint{2003.01700}.

\bibitem{Caletti:2021ysv}
S.~Caletti, O.~Fedkevych, S.~Marzani and D.~Reichelt,
\newblock \emph{{Tagging the initial-state gluon}},
\newblock Eur. Phys. J. C \textbf{81}(9), 844 (2021),
\newblock \doi{10.1140/epjc/s10052-021-09648-x},
\newblock \eprint{2108.10024}.

\bibitem{Dokshitzer:1991fd}
Y.~L. Dokshitzer, V.~A. Khoze and S.~I. Troian,
\newblock \emph{{On specific QCD properties of heavy quark fragmentation ('dead
  cone')}},
\newblock J. Phys. G \textbf{17}, 1602 (1991),
\newblock \doi{10.1088/0954-3899/17/10/023}.

\bibitem{Banfi:2004yd}
A.~Banfi, G.~P. Salam and G.~Zanderighi,
\newblock \emph{{Principles of general final-state resummation and automated
  implementation}},
\newblock JHEP \textbf{03}, 073 (2005),
\newblock \doi{10.1088/1126-6708/2005/03/073},
\newblock \eprint{hep-ph/0407286}.

\bibitem{Wang:2023azz}
X.-R. Wang and B.~Yan,
\newblock \emph{{Probing the Hgg coupling through the jet charge correlation in
  Higgs boson decay}},
\newblock Phys. Rev. D \textbf{108}(5), 056010 (2023),
\newblock \doi{10.1103/PhysRevD.108.056010},
\newblock \eprint{2302.02084}.

\bibitem{Perez:2015lra}
G.~Perez, Y.~Soreq, E.~Stamou and K.~Tobioka,
\newblock \emph{{Prospects for measuring the Higgs boson coupling to light
  quarks}},
\newblock Phys. Rev. D \textbf{93}(1), 013001 (2016),
\newblock \doi{10.1103/PhysRevD.93.013001},
\newblock \eprint{1505.06689}.

\bibitem{ATLAS:2017gko}
M.~Aaboud \emph{et~al.},
\newblock \emph{{Search for exclusive Higgs and $Z$ boson decays to
  $\phi\gamma$ and $\rho\gamma$ with the ATLAS detector}},
\newblock JHEP \textbf{07}, 127 (2018),
\newblock \doi{10.1007/JHEP07(2018)127},
\newblock \eprint{1712.02758}.

\bibitem{Chahal:2022rid}
G.~S. Chahal and F.~Krauss,
\newblock \emph{{Cluster Hadronisation in Sherpa}},
\newblock SciPost Phys. \textbf{13}(2), 019 (2022),
\newblock \doi{10.21468/SciPostPhys.13.2.019},
\newblock \eprint{2203.11385}.

\bibitem{Sherpa:2019gpd}
E.~Bothmann \emph{et~al.},
\newblock \emph{{Event Generation with Sherpa 2.2}},
\newblock SciPost Phys. \textbf{7}(3), 034 (2019),
\newblock \doi{10.21468/SciPostPhys.7.3.034},
\newblock \eprint{1905.09127}.

\bibitem{ATLAS:2019kwg}
M.~Aaboud \emph{et~al.},
\newblock \emph{{Measurement of jet-substructure observables in top quark, $W$
  boson and light jet production in proton-proton collisions at $\sqrt{s}=13$
  TeV with the ATLAS detector}},
\newblock JHEP \textbf{08}, 033 (2019),
\newblock \doi{10.1007/JHEP08(2019)033},
\newblock \eprint{1903.02942}.

\bibitem{CMS:2021iwu}
A.~Tumasyan \emph{et~al.},
\newblock \emph{{Study of quark and gluon jet substructure in Z+jet and dijet
  events from pp collisions}},
\newblock JHEP \textbf{01}, 188 (2022),
\newblock \doi{10.1007/JHEP01(2022)188},
\newblock \eprint{2109.03340}.

\bibitem{ALICE:2021njq}
S.~Acharya \emph{et~al.},
\newblock \emph{{Measurements of the groomed and ungroomed jet angularities in
  pp collisions at $ \sqrt{s} $ = 5.02 TeV}},
\newblock JHEP \textbf{05}, 061 (2022),
\newblock \doi{10.1007/JHEP05(2022)061},
\newblock \eprint{2107.11303}.

\bibitem{Caletti:2021oor}
S.~Caletti, O.~Fedkevych, S.~Marzani, D.~Reichelt, S.~Schumann, G.~Soyez and
  V.~Theeuwes,
\newblock \emph{{Jet angularities in Z+jet production at the LHC}},
\newblock JHEP \textbf{07}, 076 (2021),
\newblock \doi{10.1007/JHEP07(2021)076},
\newblock \eprint{2104.06920}.

\bibitem{Reichelt:2021svh}
D.~Reichelt, S.~Caletti, O.~Fedkevych, S.~Marzani, S.~Schumann and G.~Soyez,
\newblock \emph{{Phenomenology of jet angularities at the LHC}},
\newblock JHEP \textbf{03}, 131 (2022),
\newblock \doi{10.1007/JHEP03(2022)131},
\newblock \eprint{2112.09545}.

\bibitem{Gras:2017jty}
P.~Gras, S.~H\"oche, D.~Kar, A.~Larkoski, L.~L\"onnblad, S.~Pl\"atzer,
  A.~Si\'odmok, P.~Skands, G.~Soyez and J.~Thaler,
\newblock \emph{{Systematics of quark/gluon tagging}},
\newblock JHEP \textbf{07}, 091 (2017),
\newblock \doi{10.1007/JHEP07(2017)091},
\newblock \eprint{1704.03878}.

\bibitem{Larkoski:2014wba}
A.~J. Larkoski, S.~Marzani, G.~Soyez and J.~Thaler,
\newblock \emph{{Soft Drop}},
\newblock JHEP \textbf{05}, 146 (2014),
\newblock \doi{10.1007/JHEP05(2014)146},
\newblock \eprint{1402.2657}.

\bibitem{Baron:2018nfz}
J.~Baron, S.~Marzani and V.~Theeuwes,
\newblock \emph{{Soft-Drop Thrust}},
\newblock JHEP \textbf{08}, 105 (2018),
\newblock \doi{10.1007/JHEP08(2018)105},
\newblock [Erratum: JHEP 05, 056 (2019)],
\newblock \eprint{1803.04719}.

\bibitem{Marzani:2019evv}
S.~Marzani, D.~Reichelt, S.~Schumann, G.~Soyez and V.~Theeuwes,
\newblock \emph{{Fitting the Strong Coupling Constant with Soft-Drop Thrust}},
\newblock JHEP \textbf{11}, 179 (2019),
\newblock \doi{10.1007/JHEP11(2019)179},
\newblock \eprint{1906.10504}.

\bibitem{Baron:2020xoi}
J.~Baron, D.~Reichelt, S.~Schumann, N.~Schwanemann and V.~Theeuwes,
\newblock \emph{{Soft-drop grooming for hadronic event shapes}},
\newblock JHEP \textbf{07}, 142 (2021),
\newblock \doi{10.1007/JHEP07(2021)142},
\newblock \eprint{2012.09574}.

\bibitem{Dokshitzer:1997in}
Y.~L. Dokshitzer, G.~D. Leder, S.~Moretti and B.~R. Webber,
\newblock \emph{{Better jet clustering algorithms}},
\newblock JHEP \textbf{08}, 001 (1997),
\newblock \doi{10.1088/1126-6708/1997/08/001},
\newblock \eprint{hep-ph/9707323}.

\bibitem{Wobisch:1998wt}
M.~Wobisch and T.~Wengler,
\newblock \emph{{Hadronization corrections to jet cross-sections in deep
  inelastic scattering}},
\newblock In \emph{{Workshop on Monte Carlo Generators for HERA Physics
  (Plenary Starting Meeting)}}, pp. 270--279 (1998), \eprint{hep-ph/9907280}.

\bibitem{Butterworth:2008iy}
J.~M. Butterworth, A.~R. Davison, M.~Rubin and G.~P. Salam,
\newblock \emph{{Jet substructure as a new Higgs search channel at the LHC}},
\newblock Phys. Rev. Lett. \textbf{100}, 242001 (2008),
\newblock \doi{10.1103/PhysRevLett.100.242001},
\newblock \eprint{0802.2470}.

\bibitem{Dasgupta:2013ihk}
M.~Dasgupta, A.~Fregoso, S.~Marzani and G.~P. Salam,
\newblock \emph{{Towards an understanding of jet substructure}},
\newblock JHEP \textbf{09}, 029 (2013),
\newblock \doi{10.1007/JHEP09(2013)029},
\newblock \eprint{1307.0007}.

\bibitem{Sherpa3.0.beta}
\emph{\normalfont{The \sherpa-3.0.beta code can be obtained from:}
  \url{https://sherpa-team.gitlab.io/changelog.html}}.

\bibitem{Schumann:2007mg}
S.~Schumann and F.~Krauss,
\newblock \emph{{A Parton shower algorithm based on Catani-Seymour dipole
  factorisation}},
\newblock JHEP \textbf{03}, 038 (2008),
\newblock \doi{10.1088/1126-6708/2008/03/038},
\newblock \eprint{0709.1027}.

\bibitem{Gleisberg:2008ta}
T.~Gleisberg, S.~H{\"o}che, F.~Krauss, M.~Sch{\"o}nherr, S.~Schumann,
  F.~Siegert and J.~Winter,
\newblock \emph{{Event generation with SHERPA 1.1}},
\newblock JHEP \textbf{02}, 007 (2009),
\newblock \doi{10.1088/1126-6708/2009/02/007},
\newblock \eprint{0811.4622}.

\bibitem{Bierlich:2019rhm}
C.~Bierlich \emph{et~al.},
\newblock \emph{{Robust Independent Validation of Experiment and Theory: Rivet
  version 3}},
\newblock SciPost Phys. \textbf{8}, 026 (2020),
\newblock \doi{10.21468/SciPostPhys.8.2.026},
\newblock \eprint{1912.05451}.

\bibitem{Butterworth:2016sqg}
J.~M. Butterworth, D.~Grellscheid, M.~Kr\"amer, B.~Sarrazin and D.~Yallup,
\newblock \emph{{Constraining new physics with collider measurements of
  Standard Model signatures}},
\newblock JHEP \textbf{03}, 078 (2017),
\newblock \doi{10.1007/JHEP03(2017)078},
\newblock \eprint{1606.05296}.

\bibitem{Buckley:2021neu}
A.~Buckley \emph{et~al.},
\newblock \emph{{Testing new physics models with global comparisons to collider
  measurements: the Contur toolkit}},
\newblock SciPost Phys. Core \textbf{4}, 013 (2021),
\newblock \doi{10.21468/SciPostPhysCore.4.2.013},
\newblock \eprint{2102.04377}.

\bibitem{Ellis:1975ap}
J.~R. Ellis, M.~K. Gaillard and D.~V. Nanopoulos,
\newblock \emph{{A Phenomenological Profile of the Higgs Boson}},
\newblock Nucl. Phys. B \textbf{106}, 292 (1976),
\newblock \doi{10.1016/0550-3213(76)90382-5}.

\bibitem{Shifman:1979eb}
M.~A. Shifman, A.~I. Vainshtein, M.~B. Voloshin and V.~I. Zakharov,
\newblock \emph{{Low-Energy Theorems for Higgs Boson Couplings to Photons}},
\newblock Sov. J. Nucl. Phys. \textbf{30}, 711 (1979).

\bibitem{Kniehl:1995tn}
B.~A. Kniehl and M.~Spira,
\newblock \emph{{Low-energy theorems in Higgs physics}},
\newblock Z. Phys. C \textbf{69}, 77 (1995),
\newblock \doi{10.1007/s002880050007},
\newblock \eprint{hep-ph/9505225}.

\bibitem{LHCHiggsCrossSectionWorkingGroup:2016ypw}
D.~de~Florian \emph{et~al.},
\newblock \emph{{Handbook of LHC Higgs Cross Sections: 4. Deciphering the
  Nature of the Higgs Sector}} \textbf{2/2017} (2016),
\newblock \doi{10.23731/CYRM-2017-002},
\newblock \eprint{1610.07922}.

\bibitem{Albert:2022mpk}
A.~Albert \emph{et~al.},
\newblock \emph{{Strange quark as a probe for new physics in the Higgs
  sector}},
\newblock In \emph{{Snowmass 2021}} (2022), \eprint{2203.07535}.

\bibitem{Hoeche:2012yf}
S.~H{\"o}che, F.~Krauss, M.~Sch{\"o}nherr and F.~Siegert,
\newblock \emph{{QCD matrix elements + parton showers: The NLO case}},
\newblock JHEP \textbf{04}, 027 (2013),
\newblock \doi{10.1007/JHEP04(2013)027},
\newblock \eprint{1207.5030}.

\bibitem{Gleisberg:2008fv}
T.~Gleisberg and S.~H{\"o}che,
\newblock \emph{{Comix, a new matrix element generator}},
\newblock JHEP \textbf{12}, 039 (2008),
\newblock \doi{10.1088/1126-6708/2008/12/039},
\newblock \eprint{0808.3674}.

\bibitem{Krauss:2001iv}
F.~Krauss, R.~Kuhn and G.~Soff,
\newblock \emph{{AMEGIC++ 1.0: A Matrix element generator in C++}},
\newblock JHEP \textbf{02}, 044 (2002),
\newblock \doi{10.1088/1126-6708/2002/02/044},
\newblock \eprint{hep-ph/0109036}.

\bibitem{Buccioni:2019sur}
F.~Buccioni, J.-N. Lang, J.~M. Lindert, P.~Maierh\"ofer, S.~Pozzorini, H.~Zhang
  and M.~F. Zoller,
\newblock \emph{{OpenLoops 2}},
\newblock Eur. Phys. J. C \textbf{79}(10), 866 (2019),
\newblock \doi{10.1140/epjc/s10052-019-7306-2},
\newblock \eprint{1907.13071}.

\bibitem{Hoeche:2009rj}
S.~H{\"o}che, F.~Krauss, S.~Schumann and F.~Siegert,
\newblock \emph{{QCD matrix elements and truncated showers}},
\newblock JHEP \textbf{05}, 053 (2009),
\newblock \doi{10.1088/1126-6708/2009/05/053},
\newblock \eprint{0903.1219}.

\bibitem{Krishnamoorthy:2021nwv}
M.~Krishnamoorthy, H.~Schulz, X.~Ju, W.~Wang, S.~Leyffer, Z.~Marshall,
  S.~Mrenna, J.~M\"uller and J.~B. Kowalkowski,
\newblock \emph{{Apprentice for Event Generator Tuning}},
\newblock EPJ Web Conf. \textbf{251}, 03060 (2021),
\newblock \doi{10.1051/epjconf/202125103060},
\newblock \eprint{2103.05748}.

\bibitem{Winter:2003tt}
J.-C. Winter, F.~Krauss and G.~Soff,
\newblock \emph{{A Modified cluster hadronization model}},
\newblock Eur. Phys. J. C \textbf{36}, 381 (2004),
\newblock \doi{10.1140/epjc/s2004-01960-8},
\newblock \eprint{hep-ph/0311085}.

\bibitem{ALEPH:1991ldi}
D.~Decamp \emph{et~al.},
\newblock \emph{{Measurement of the charged particle multiplicity distribution
  in hadronic Z decays}},
\newblock Phys. Lett. B \textbf{273}, 181 (1991),
\newblock \doi{10.1016/0370-2693(91)90575-B}.

\bibitem{ALEPH:2003obs}
A.~Heister \emph{et~al.},
\newblock \emph{{Studies of QCD at e+ e- centre-of-mass energies between 91-GeV
  and 209-GeV}},
\newblock Eur. Phys. J. C \textbf{35}, 457 (2004),
\newblock \doi{10.1140/epjc/s2004-01891-4}.

\bibitem{DELPHI:1996sen}
P.~Abreu \emph{et~al.},
\newblock \emph{{Tuning and test of fragmentation models based on identified
  particles and precision event shape data}},
\newblock Z. Phys. C \textbf{73}, 11 (1996),
\newblock \doi{10.1007/s002880050295}.

\bibitem{ALEPH:2001pfo}
A.~Heister \emph{et~al.},
\newblock \emph{{Study of the fragmentation of b quarks into B mesons at the Z
  peak}},
\newblock Phys. Lett. B \textbf{512}, 30 (2001),
\newblock \doi{10.1016/S0370-2693(01)00690-6},
\newblock \eprint{hep-ex/0106051}.

\bibitem{OPAL:2002plk}
G.~Abbiendi \emph{et~al.},
\newblock \emph{{Inclusive analysis of the b quark fragmentation function in Z
  decays at LEP}},
\newblock Eur. Phys. J. C \textbf{29}, 463 (2003),
\newblock \doi{10.1140/epjc/s2003-01229-x},
\newblock \eprint{hep-ex/0210031}.

\bibitem{JADE:1999zar}
P.~Pfeifenschneider \emph{et~al.},
\newblock \emph{{QCD analyses and determinations of alpha(s) in e+ e-
  annihilation at energies between 35-GeV and 189-GeV}},
\newblock Eur. Phys. J. C \textbf{17}, 19 (2000),
\newblock \doi{10.1007/s100520000432},
\newblock \eprint{hep-ex/0001055}.

\bibitem{ParticleDataGroup:2008zun}
C.~Amsler \emph{et~al.},
\newblock \emph{{Review of Particle Physics}},
\newblock Phys. Lett. B \textbf{667}, 1 (2008),
\newblock \doi{10.1016/j.physletb.2008.07.018}.

\bibitem{Gerwick:2014gya}
E.~Gerwick, S.~H{\"o}che, S.~Marzani and S.~Schumann,
\newblock \emph{{Soft evolution of multi-jet final states}},
\newblock JHEP \textbf{02}, 106 (2015),
\newblock \doi{10.1007/JHEP02(2015)106},
\newblock \eprint{1411.7325}.

\bibitem{Baberuxki:2019ifp}
N.~Baberuxki, C.~T. Preuss, D.~Reichelt and S.~Schumann,
\newblock \emph{{Resummed predictions for jet-resolution scales in multijet
  production in e$^{+}$e$^{-}$ annihilation}},
\newblock JHEP \textbf{04}, 112 (2020),
\newblock \doi{10.1007/JHEP04(2020)112},
\newblock \eprint{1912.09396}.

\bibitem{Banfi:2014sua}
A.~Banfi, H.~McAslan, P.~F. Monni and G.~Zanderighi,
\newblock \emph{{A general method for the resummation of event-shape
  distributions in $e^{+} e^{-}$ annihilation}},
\newblock JHEP \textbf{05}, 102 (2015),
\newblock \doi{10.1007/JHEP05(2015)102},
\newblock \eprint{1412.2126}.

\bibitem{Banfi:2016zlc}
A.~Banfi, H.~McAslan, P.~F. Monni and G.~Zanderighi,
\newblock \emph{{The two-jet rate in $e^+e^-$ at
  next-to-next-to-leading-logarithmic order}},
\newblock Phys. Rev. Lett. \textbf{117}(17), 172001 (2016),
\newblock \doi{10.1103/PhysRevLett.117.172001},
\newblock \eprint{1607.03111}.

\bibitem{Abbate:2010xh}
R.~Abbate, M.~Fickinger, A.~H. Hoang, V.~Mateu and I.~W. Stewart,
\newblock \emph{{Thrust at $N^{3}LL$ with Power Corrections and a Precision
  Global Fit for $\alpha_{s}(mZ)$}},
\newblock Phys. Rev. D \textbf{83}, 074021 (2011),
\newblock \doi{10.1103/PhysRevD.83.074021},
\newblock \eprint{1006.3080}.

\bibitem{Hoang:2014wka}
A.~H. Hoang, D.~W. Kolodrubetz, V.~Mateu and I.~W. Stewart,
\newblock \emph{{$C$-parameter distribution at N$^3$LL' including power
  corrections}},
\newblock Phys. Rev. D \textbf{91}(9), 094017 (2015),
\newblock \doi{10.1103/PhysRevD.91.094017},
\newblock \eprint{1411.6633}.

\bibitem{Ellis:2010rwa}
S.~D. Ellis, C.~K. Vermilion, J.~R. Walsh, A.~Hornig and C.~Lee,
\newblock \emph{{Jet Shapes and Jet Algorithms in SCET}},
\newblock JHEP \textbf{11}, 101 (2010),
\newblock \doi{10.1007/JHEP11(2010)101},
\newblock \eprint{1001.0014}.

\bibitem{Larkoski:2013eya}
A.~J. Larkoski, G.~P. Salam and J.~Thaler,
\newblock \emph{{Energy Correlation Functions for Jet Substructure}},
\newblock JHEP \textbf{06}, 108 (2013),
\newblock \doi{10.1007/JHEP06(2013)108},
\newblock \eprint{1305.0007}.

\bibitem{Larkoski:2014uqa}
A.~J. Larkoski, D.~Neill and J.~Thaler,
\newblock \emph{{Jet Shapes with the Broadening Axis}},
\newblock JHEP \textbf{04}, 017 (2014),
\newblock \doi{10.1007/JHEP04(2014)017},
\newblock \eprint{1401.2158}.

\bibitem{Hornig:2016ahz}
A.~Hornig, Y.~Makris and T.~Mehen,
\newblock \emph{{Jet Shapes in Dijet Events at the LHC in SCET}},
\newblock JHEP \textbf{04}, 097 (2016),
\newblock \doi{10.1007/JHEP04(2016)097},
\newblock \eprint{1601.01319}.

\bibitem{Kang:2018qra}
Z.-B. Kang, K.~Lee and F.~Ringer,
\newblock \emph{{Jet angularity measurements for single inclusive jet
  production}},
\newblock JHEP \textbf{04}, 110 (2018),
\newblock \doi{10.1007/JHEP04(2018)110},
\newblock \eprint{1801.00790}.

\bibitem{Fickinger:2016rfd}
M.~Fickinger, S.~Fleming, C.~Kim and E.~Mereghetti,
\newblock \emph{{Effective field theory approach to heavy quark
  fragmentation}},
\newblock JHEP \textbf{11}, 095 (2016),
\newblock \doi{10.1007/JHEP11(2016)095},
\newblock \eprint{1606.07737}.

\bibitem{Gaggero:2022hmv}
D.~Gaggero, A.~Ghira, S.~Marzani and G.~Ridolfi,
\newblock \emph{{Soft logarithms in processes with heavy quarks}},
\newblock JHEP \textbf{09}, 058 (2022),
\newblock \doi{10.1007/JHEP09(2022)058},
\newblock \eprint{2207.13567}.

\bibitem{Ghira:2023bxr}
A.~Ghira, S.~Marzani and G.~Ridolfi,
\newblock \emph{{A consistent resummation of mass and soft logarithms in
  processes with heavy flavours}}  (2023),
\newblock \eprint{2309.06139}.

\bibitem{ALEPH:1999syy}
R.~Barate \emph{et~al.},
\newblock \emph{{Study of charm production in Z decays}},
\newblock Eur. Phys. J. C \textbf{16}, 597 (2000),
\newblock \doi{10.1007/s100520000421},
\newblock \eprint{hep-ex/9909032}.

\end{thebibliography}

\end{document}